\documentclass[12pt,onecolumn,draftcls]{IEEEtran}

\usepackage{setspace}
\usepackage{verbatim}
\usepackage{graphicx}
\usepackage{cite}
\usepackage{algorithmic}
\usepackage{algorithm}
\usepackage{subfigure}
\usepackage{amssymb}

\usepackage[cmex10]{amsmath}
\usepackage{url}
\usepackage{epstopdf}

\newtheorem{thm}{Theorem} 
\newtheorem{lem}[thm]{Lemma}

\newtheorem{cor}[thm]{Corollary}

\newtheorem{defn}{Definition} 

\newtheorem{remk}{Remark} 

\begin{document}
%
\title{Quantized Iterative Message Passing Decoders with Low Error Floor for LDPC Codes}

\author{Xiaojie~Zhang\IEEEmembership{,~Member,~IEEE}%
        ~and~Paul~H.~Siegel\IEEEmembership{,~Fellow,~IEEE}
\thanks{This research was supported in part by the Center for Magnetic Recording Research at University of California, San Diego and by the National Science Foundation under Grants CCF-0829865 and CCF-1116739, and University of California Lab Fees Research Program, Award No. 09-LR-06-118620-SIEP.}%
\thanks{The material in this paper was presented in part at the IEEE International Symposium on Information Theory, Cambridge, MA, July 1--5, 2012, and IEEE International Conference on Signal Processing and Communication, Bangalore, India, July 22--25, 2012.}%
\thanks{Xiaojie Zhang and Paul H. Siegel are with the Department of Electrical and Computer Engineering and the Center for Magnetic Recording Research, University of California, San Diego, La Jolla, CA 92093 (email: \{ericzhang, psiegel\}@ucsd.edu)}}

\markboth{Submitted to IEEE Trans. Commun.}{}

\maketitle

\vspace{-0.75in}
\begin{abstract}
The error floor phenomenon observed with LDPC codes and their graph-based, iterative, message-passing (MP) decoders is commonly attributed to the existence of error-prone substructures -- variously referred to as near codewords, trapping sets, absorbing sets, or pseudocodewords -- in a Tanner graph representation of the code.  Many approaches have been proposed to lower the error floor by designing new LDPC codes with fewer such substructures or by modifying the decoding algorithm. Using a theoretical analysis of iterative MP decoding in an idealized trapping set scenario,  we show that a contributor to the error floors observed in the literature may be the imprecise implementation of decoding algorithms and, in particular, the message quantization rules used. We then propose a new quantization method -- ($q+1)$-bit quasi-uniform quantization -- that efficiently increases the dynamic range of messages, thereby overcoming a limitation of conventional quantization schemes.
Finally, we use the quasi-uniform quantizer to decode several LDPC codes that suffer from high error floors with traditional fixed-point decoder implementations. The performance simulation results provide evidence that the proposed quantization scheme can, for a wide variety of codes, significantly lower error floors with minimal increase in decoder complexity.

\end{abstract}


\IEEEpeerreviewmaketitle

\section{Introduction}
\label{sec:intro}
\IEEEPARstart{T}{he} outstanding performance of low-density parity-check (LDPC) codes and iterative, message-passing (MP) decoding algorithms~\cite{Gallager,MacKay} has attracted considerable attention over the past decade and these techniques are being deployed in a growing number of practical applications. At high signal-to-noise ratio (SNR), however, LDPC codes and MP decoders may be subject to the error floor phenomenon, which manifests itself as an abrupt change in the slope of the error-rate curve. Since many important applications, such as data storage and high-speed digital communication, often require extremely low error rates, the study of error floors in LDPC codes remains of considerable practical, as well as theoretical, interest.

The error floor phenomenon is commonly attributed to the existence of certain error-prone substructures (EPSs) in a Tanner graph representation of the code. In the binary erasure channel (BEC), it has been shown that substructures known as \emph{stopping sets} determine the error-rate performance and the observed error floor~\cite{stoppingset}. However, for general memoryless binary-input output-symmetric (MBIOS) channels such as the binary symmetric channel (BSC) and the additive white gaussian noise channel (AWGNC), the EPSs that dominate the error floor performance have not yet been fully characterized, although some classes of EPSs have been identified and studied, such as \emph{near-codewords}~\cite{nearcodeword}, \emph{trapping sets}~\cite{Richardson_ef}, \emph{absorbing sets}~\cite{absorbingsets}, and \emph{pseudocodewords}~\cite{pseudocodeword}.

One common way to improve the error floor performance of LDPC codes has been to redesign the codes to have Tanner graphs with large girth and without problematic EPSs which usually consist of small number of variable nodes~\cite{protograph,hdr,Vasic_ts}. However, for LDPC codes that have been standardized, approaches are needed that do not modify the codes. In the literature, many modifications to the iterative MP decoding algorithms have been proposed in order to improve high SNR performance, such as averaged decoders \cite{avgBP}, reordered decoders \cite{reorderBP,informedBP}, and decoders with post processing \cite{postBP,postFAS,postReliability,postAugment,han_ryan}.
In~\cite{avgBP}, the authors noticed that the emergence of errors in EPSs is heuristically related to a sudden magnitude change in the values of certain variable nodes (VNs). Hence, it was proposed to average the messages in a belief-propagation (BP) decoder over several iterations to avoid such sudden changes and therefore slow down the convergence rate for variable nodes in a trapping set and decrease the frequency of trapping set errors.
Another heuristic approach is to process messages based on the order of node reliabilities computed at each iteration~\cite{reorderBP}, and it was suggested that the scheduled decoders are able to resolve some standard trapping set errors \cite{informedBP}. Although these general approaches are capable of improving the average error rate performance to some extent, the resulting decoders still fail on small EPSs and their effect on the error floor is not significant.

To further improve the error floor behavior, decoders that make use of the prior knowledge of some small size EPSs have been designed to reduce the decoding failures due to such EPSs.
In~\cite{postBP} and \cite{postFAS}, the authors proposed a post-processing decoder that matches the configuration of unsatisfied check nodes (CNs) to trapping sets in a precomputed list after conventional MP decoding has failed. The size and completeness of the trapping set list directly affect the performance gain of such decoders, but to obtain a complete list of small trapping sets of a given LDPC code is generally quite computationally complex.
A symbol-selecting post-processing technique was also developed in~\cite{postAugment}. It saturates the channel messages on a set of selected variable nodes at each stage after the conventional MP algorithms fails. In \cite{han_ryan}, Han and Ryan proposed a bi-mode erasure decoder that combines several problematic check nodes into a generalized constraint processor, to which a corresponding maximum a posteriori (MAP) algorithm, such as the BCJR algorithm, is then applied. Another post-processing approach that utilizes the graph-theoretic structure of absorbing sets, proposed in \cite{postReliability}, adjusts the appropriate messages in the iterative MP decoding once the decoder enters and remains in the absorbing set of interest.

All the above approaches either change the message update rules of MP decoders or require extra processing steps after conventional MP decoding fails, both of which increase the decoding complexity relative to the original iterative MP algorithms. Moreover, the post-processing approaches that require prior knowledge of the set of EPSs causing the error floor are only effective when applied to LDPC codes whose EPSs have been carefully studied.

In fixed-point implementation of iterative MP decoding, efforts have also been made to improve the error-rate performance in the waterfall region and/or error-floor region by optimizing parameters of uniform quantization \cite{Zhao,fixedZhang,ZhangThesis,Tong}. In \cite{Zhao},
Zhao \emph{et al.} studied the effect of message clipping and uniform quantization on the performance of the min-sum decoder  in the waterfall region and heuristically optimized the number of quantization bits and the quantization step size for selected LDPC codes.
In \cite{Tong}, a dual-mode adaptive uniform quantization scheme was proposed to better approximate the log-tanh function used in  sum-product algorithm (SPA) decoding. Specifically, for magnitudes less than 1, all quantization bits were used to represent the fractional part; for magnitudes greater than or equal to 1, all bits were dedicated to the representation of the integer part.
In \cite{fixedZhang,ZhangThesis}, Zhang \emph{et al.} proposed a conceptually similar idea to increase precision in the quantization of the log-tanh function. Uniform quantization was applied to messages generated by both variable nodes and check nodes, but the quantization step sizes used in the two cases were separately optimized.
We note, however, that none of these modified quantization schemes were primarily intended to significantly increase the saturation level, or range, of quantized messages, and in their reported simulation results, error floors can still be clearly observed.

It has been observed that the high error floors associated with certain EPSs of some LDPC codes are closely related to the saturation level imposed on messages passed in the SPA decoder. (See, for example, \cite{Brian} and references cited therein.)
In this work, we investigate the cause of error floors in binary LDPC codes from the  perspective of the MP decoder implementation, with special attention to limitations that decrease the
dynamic range of messages passed during decoding. We show that, under certain idealized assumptions, the EPSs which are commonly associated with high error floors of some LDPC codes will not trap iterative MP decoders and cause high error floors if
message magnitudes and the number of iterations are not limited.
Based upon an analysis of the growth rate of messages outside an EPS in an idealized scenario, we propose a novel quasi-uniform quantization method that captures the essence of messages in different ranges of reliability. The proposed quantization method has an extremely large saturation level which prevents iterative MP decoders from being trapped by an EPS. This property, to the best of our knowledge, distinguishes it from other quantization techniques for iterative MP decoding that have appeared in the literature. With the new quantization method, it is possible to have a fixed point implementation of iterative MP decoders that achieves low error floors without an additional post-processing stage or a modification of either the decoding update rules or the graphical code representation upon which the iterative MP decoder operates.
We present simulation results for min-sum decoding, SPA decoding, and some of their variants, that demonstrate a significant reduction in the error floors of four representative LDPC codes, with no increase in decoding complexity.

The remainder of the paper is organized as follows. Section~\ref{sec:ND} gives some notation and definitions used throughout the paper. In Section~\ref{sec:EF}, we analytically investigate the impact that message quantization can have on MP decoder performance and the error floor phenomenon. In Section~\ref{sec:QuanDec}, we propose an enhanced quantization method intended to overcome the limitations of traditional quantization rules. In Section~\ref{sec:nr}, we incorporate the new quantizer into SPA and min-sum decoding and, through computer simulation of several LDPC codes known for their high error floors, demonstrate the significant improvement in error-rate performance that this new quantization approach can afford. Section~\ref{sec:con} concludes the paper.

\section{Notation and Definitions}
\label{sec:ND}
The study of the phenomenon of error floors began shortly after LDPC codes were rediscovered about a decade ago. It has been shown that the EPSs known as \emph{stopping sets} cause the error floor in the binary erasure channel (BEC), and such EPSs have a clear combinatorial description. Enumeration of these structures makes it possible to accurately estimate the error floor~\cite{stoppingset}.
However, for other MBIOS channels such as the binary symmetric channel (BSC) and the additive white Gaussian noise channel (AWGNC), it is more difficult to establish the relationship between EPSs and error floors. In~\cite{nearcodeword}, it was first pointed out that the \emph{near-codewords} caused error floors in simulations of Margulis and Ramanujan-Margulis LDPC codes on the AWGNC.
The term \emph{trapping set} proposed by Richardson~\cite{Richardson_ef} is operationally defined as a subset of variable nodes (VNs) that is susceptible to errors under a certain iterative MP decoder over an MBIOS channel. Hence, this concept depends on both the channel and the decoding algorithm.
In \cite{absorbingsets}, the error floor is associated with some combinatorial substructures within the Tanner graph, named \emph{absorbing sets}, which are defined independently of the channel. The absorbing sets correspond to a particular type of near codewords or trapping sets that are stable under bit-flipping operations.
All these EPSs have been believed to be the cause of error floors, and for some LDPC codes, techniques such as importance sampling used to estimate the error floor are based on the probability of decoding failures on such EPSs \cite{Richardson_ef,ImportSamp}.
In this section, we will show that under certain idealized assumptions about the computation trees of variable nodes within a given EPS, as well as the correctness of variable node messages outside the  EPS, conventional iterative decoders that accurately represent messages will eventually correct errors supported by the EPS.

To facilitate our discussion, we define a substructure called an \emph{absolute trapping set} from a purely graph-theoretic perspective, independent of the channel and the decoder. Let $G=(V\cup C,E)$ denote the Tanner graph of a binary LDPC code with VNs $V=\{v_1,\dots,v_n\}$, CNs $C=\{c_1,\dots,c_m\}$, and edge set $E$.

\begin{defn}
\label{def_eps}
A \emph{stopping set} of size $a$ is a configuration of $a$ variable nodes such that the induced subgraph has no check nodes of degree-one.
An $(a, b)$ \emph{trapping set} is a configuration of $a$ variable nodes, for which the induced subgraph is connected and has $b$ odd-degree check nodes. If the induced subgraph of an $(a, b)$ trapping set does not contain a stopping set
, it is called an \emph{absolute trapping set}.
\end{defn}

In the literature, all trapping sets of interest that contribute to the error floor of an LDPC code are of size smaller than the minimum stopping set size of the code, since otherwise the stopping sets would be the dominant contributor to the error floor~\cite{stoppingset}.
Note that the requirement that an absolute trapping set contain no stopping set also implies that it must have at lease one degree-one check node.
As we will discuss later in this section, these degree-one check nodes are essential because they are able to pass correct extrinsic messages into the trapping set. To the best of our knowledge, almost all trapping sets of interest in the literature are absolute trapping sets. For example, both of the well-known (5,3) trapping sets in the Tanner code of length 155, the notorious (12,4) trapping sets in the (2640,1320) Margulis code, and the (5,5) trapping set in some codes of variable-degree five are all absolute trapping sets.
Unless otherwise indicated, all trapping sets referred to in this paper are absolute trapping sets, as well.

In analogy to the definition of \emph{computation tree} in \cite{comptree}, we define a \emph{k-iteration computation tree} as follows.

\begin{defn}
\label{def_ct}
A \emph{k-iteration computation tree} $T_k(v)$ for an iterative decoder in the Tanner graph $G$ is a tree graph constructed by choosing variable node $v\in V$ as its root and then recursively adding edges and leaf nodes to the tree that participate in the iterative message-passing decoding during $k$ iterations. To each vertex that is created in $T_k(v)$, we associate the corresponding node update function in $G$.
\end{defn}

Let $S$ be the induced subgraph of an $(a,b)$ trapping set contained in $G$, with VN set $V_S\subseteq V$ and CN set $C_S\subseteq C$. Let set $C_1\subseteq C_S$ be the set of degree-one CNs in the subgraph $S$, and let set $V_1\subseteq V_S$ be the set of neighboring VNs of CNs in $C_1$.
We refer to a message on an edge adjacent to VN $v$ as a \emph{correct} message if its sign reflects the correct value of $v$, and as an \emph{incorrect} message, otherwise.
Let $D(u)$ be the set of all descendants of the vertex $u$ in a given computation tree.

\begin{defn}
\label{def_separation}
Given a Tanner graph $G$ and an induced subgraph $S$ of a trapping set, a variable node $v\in V_1$ is said to be \emph{$k$-separated} if, for at least one of its neighboring degree-one check node $c\in C_1$, no variable node $v'\in V_S$ belongs to $D(c) \subset T_k(v)$. If every $v\in V_1$ is $k$-separated, the induced subgraph $S$ is said to satisfy the \emph{$k$-separation assumption}.
\end{defn}

\begin{figure*}[!t]
\centerline
{\subfigure[$(4,4)$ trapping set and part of its neighboring nodes]{\includegraphics[width=0.35\linewidth]{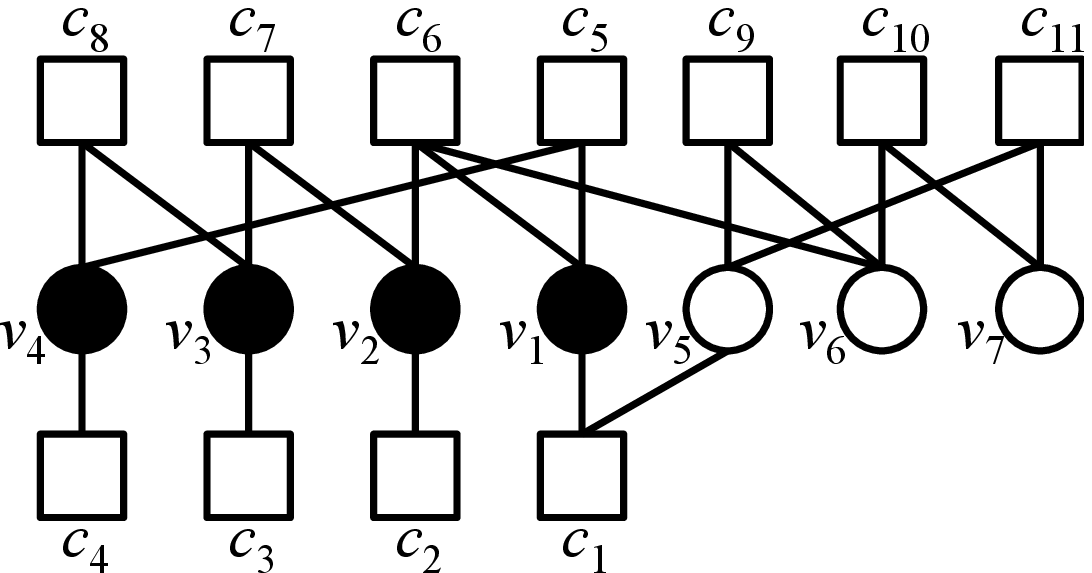}
\label{subfig_TS44}}
\hfil
\subfigure[Computation tree with root $v_1$]{\includegraphics[width=0.3\linewidth]{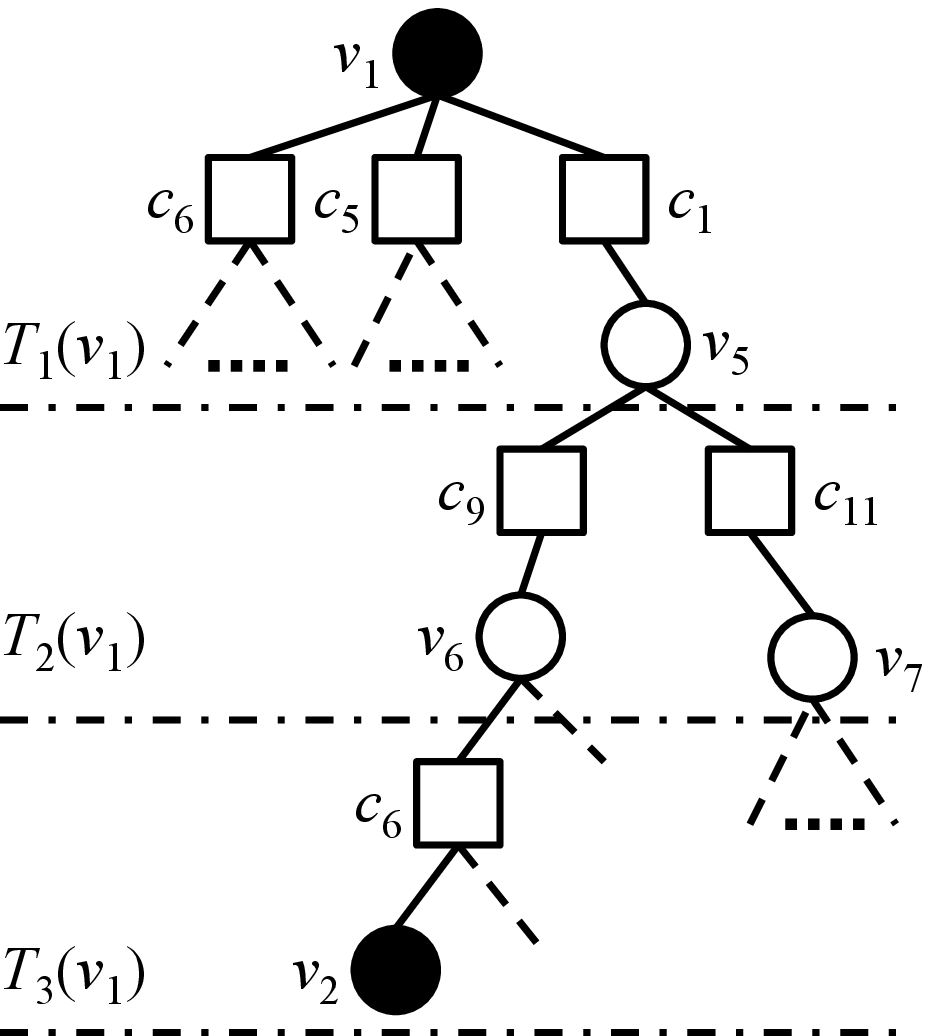}
\label{subfig_comptree}}}%
\caption{Example of a $(4,4)$ trapping set and its corresponding computation tree.}
\label{fig_tree}
\end{figure*}

In Fig.~\ref{subfig_TS44}, we show the graph of a $(4,4)$ trapping set and some of its neighboring nodes. The set of VNs in the trapping set is $V_S=\{v_1,v_2,v_3,v_4\}$, represented as solid black circles. The set of CNs in the trapping set is $C_S=\{c_i\}$, $1\leq i\leq8$. In this trapping set, every VN has a neighboring degree-one CN, i.e., $V_1=V_S$, and $C_1=\{c_1,c_2,c_3,c_4\}$.
For example, the 3-iteration computation tree of VN $v_1$ is shown in Fig.~\ref{subfig_comptree}. It can be verified from this computation tree that $v_1$ is 2-separated but not 3-separated, because $v_2\in V_S$ is a descendant of $c_1$ in $T_3(v_1)$, but not in $T_2(v_1)$. It is worth noting that whether or not a trapping set satisfies the $k$-separation assumption depends on the Tanner graph outside the trapping set, not the trapping set itself.

We want to point out that the $k$-separation assumption is much weaker than the isolation assumption in \cite{isolation}. The separation assumption here only applies to the VNs that have neighboring degree-one CNs in the induced subgraph $S$, and these neighboring degree-one CNs do not have any VNs from the trapping set as their descendants in the corresponding $k$-iteration computation tree. With the separation assumption, the descendants of $c\in C_1$ are separated from all the nodes in the trapping set, meaning that the incorrect messages passed in the trapping set do not affect the extrinsic messages sent towards $c$ in the computation tree.

\section{Error Floors of LDPC Codes}
\label{sec:EF}

\subsection{Trapping Sets and Min-Sum Decoding}

To get further insight into the connection between trapping sets and decoding failures of iterative MP decoders, we first consider a simple iterative MP decoder, the min-sum (MS) decoder, which can be viewed as a simple approximation of the sum-product algorithm.
We now briefly recall the VN and CN update rules of min-sum decoding.

A VN $v_i$ receives an input message $L^{ch}_i$ from the channel, typically the log-likelihood ratio (LLR) of the corresponding channel output, defined as follows
\begin{equation}
\label{LLR}
L^{ch}_i = \log \left( {\frac{\Pr \left( {\left. {R_i = r_i} \right|{c_i} = 0} \right)}{\Pr \left( {\left. {R_i = r_i} \right|{c_i} = 1} \right)}} \right),
\end{equation}
where $c_i\in\{0,1\}$ is the code bit and $r_i$ is the corresponding received symbol.

Denote by $L_{i\rightarrow j}$ and $L_{j\rightarrow i}$ the messages sent from $v_i$ to $c_j$ and from $c_j$ to $v_i$, respectively, and denote by $N(k)$ the set of neighboring nodes of VN $v_k$ (or CN $c_k$). Then, the message sent from $v_i$ to $c_j$ in min-sum decoding is given by
\begin{equation}
\label{eq_MS_VN}
L_{i\rightarrow j} = L^{ch}_i + \sum\limits_{j'\in N(i)\setminus j} L_{j'\rightarrow i}~,
\end{equation}
and the message from CN $j$ to VN $i$ is computed as
\begin{equation}
\label{eq_MS_CN}
L_{j\rightarrow i} = \left[\prod\limits_{i'\in N(j)\setminus i}\text{sign}(L_{i'\rightarrow j})\right]\cdot \min\limits_{i'\in N(j)\setminus i} |L_{i'\rightarrow j}|.
\end{equation}
In the initialization step, we set $L_{i\rightarrow j} = L^{ch}_i$.
It can been seen from \eqref{eq_MS_VN} and \eqref{eq_MS_CN} that the min-sum decoding algorithm is insensitive to linear scaling, meaning that linearly scaling all input messages from the channel would not affect the decoding performance.

For the MS decoder, we can show that a trapping set does not cause decoding failure if its induced subgraph in the Tanner graph satisfies certain criteria.




\begin{thm}\label{Thm_MS}
Let $G$ be the Tanner graph of a variable-regular LDPC code that contains a subgraph $S$ induced by a trapping set.
Assume that the channel is either a BSC or an AWGNC, and that the messages from the channel to all VNs outside $S$ are correct.
If $S$ satisfies the $k$-separation assumption for sufficiently large $k$, then the corresponding min-sum decoder will successfully correct all erroneous VNs in $S$.
\end{thm}
\begin{IEEEproof}
See Appendix~\ref{Appendix_Thm_MS}.
\end{IEEEproof}

In general, the error-rate performance of MS decoding is not as good as that of SPA decoding. However, there are several quite simple but effective ways to adjust the CN update rule of MS decoding to get comparable performance to SPA decoding.
One method is \emph{attenuated-min-sum} (AMS) decoding~\cite{MinSum_vari}, where the magnitudes of messages are attenuated at CNs. The corresponding CN update rule of AMS is as follows
\begin{equation}
\label{eq_AMS_CN}
L_{j\rightarrow i} = \left[\prod\limits_{i'\in N(j)\setminus i}\text{sign}(L_{i'\rightarrow j})\right]\cdot \alpha \cdot \min\limits_{i'\in N(j)\setminus i} |L_{i'\rightarrow j}|,
\end{equation}
where $0<\alpha<1$ is the attenuation factor, which can be a fixed constant or adaptively adjusted.
Another way to improve the error-rate performance of MS decoding is \emph{offset-min-sum} (OMS) decoding, which applies an offset to reduce the magnitudes of CN output messages. The resulting CN update equation is
\begin{equation}
\label{eq_OMS_CN}
L_{j\rightarrow i} = \left[\prod\limits_{i'\in N(j)\setminus i}\text{sign}(L_{i'\rightarrow j})\right]\cdot \max\{\min\limits_{i'\in N(j)\setminus i} |L_{i'\rightarrow j}|-\beta,~0\},
\end{equation}
where $\beta>0$ is the offset which, like the attenuation factor, can be a fixed constant or adaptively adjusted.
In some implementations, for additional simplicity, the attenuation factor or offset is set to be the same fixed constant for all CNs and all iterations~\cite{MinSum_vari}.

Theorem~\ref{Thm_MS} can be extended to both AMS and OMS decoding, where we assume that, in each iteration, all CNs use the same attenuation factor $\alpha$ in AMS or the same offset $\beta$ in OMS.

\begin{cor}\label{Cor_MMS}
Let $G$ be the Tanner graph of a variable-regular LDPC code that contains a subgraph $S$ induced by a trapping set.
Assume that the channel is either a BSC or an AWGNC, and that the messages from the channel to all VNs outside $S$ are correct.
If $S$ satisfies the $k$-separation assumption for sufficiently large $k$, then the both AMS and OMS decoder will successfully correct all erroneous VNs in $S$.
\end{cor}
\begin{IEEEproof}
See Appendix~\ref{Appendix_Cor_MMS}.
\end{IEEEproof}


As shown in Appendix~\ref{Appendix_Cor_MMS}, the extension to AMS decoding follows easily from Theorem~\ref{Thm_MS}.
On the other hand, the proof of the extension to the OMS decoder makes use of ideas introduced in the analysis of SPA decoding in the next subsection.

\subsection{Trapping Sets and Sum-Product Algorithm Decoding}
\label{subsec:spa}

In this subsection, we further extend Theorem~\ref{Thm_MS} to sum-product algorithm  decoding.
The optimality criterion in the design of the SPA decoder is symbol-wise maximum a \emph{posteriori} probability (MAP), and it is an optimal symbol-wise decoder on Tanner graphs without cycles.

In SPA decoding, VN nodes take log-likelihood ratios of received information from the channel as initial input messages. The VN update rule is the same as that of MS decoding described in \eqref{eq_MS_VN}, which involves the summation of all incoming extrinsic messages. In the CN update rule of SPA decoding, the message sent from CN $j$ to VN $i$ is computed as
\begin{equation}
\label{eq_SPA_CN}
L_{j\rightarrow i} = 2{\tanh^{-1}} \left( \prod\limits_{i'\in N(j)\setminus i} {\tanh \frac{L_{i'\rightarrow j}}{2}}  \right).
\end{equation}

In practical implementations of the SPA, the following equivalent CN update rule is often used
\begin{equation}
\label{eq_SPA_CNphi}
L_{j\rightarrow i} = \left[\prod\limits_{i'\in N(j)\setminus i}\text{sign}(L_{i'\rightarrow j})\right]\cdot \phi^{-1}\left(\sum\limits_{i'\in N(j)\setminus i} \phi(|L_{i'\rightarrow j}|)\right)
\end{equation}
where $\phi(x)=-\log[\tanh(x/2)]=\log\left( (e^x+1)/(e^x-1) \right)$, $\phi^{-1}(x)=\phi(x)$, and $\phi(\infty)=0$. 
In some fixed-point implementations, in order to have better approximation, different look-up tables could be used to compute $\phi(x)$ and $\phi^{-1}(x)$~\cite{ZhangThesis}.

We note that the hyperbolic tangent function, $\tanh(x)$, has numerical saturation problems when computed with finite precision. For example, in 64-bit floating-point (in IEEE 754 standard format \cite{IEEE754}) computer implementation, it can be shown that $\tanh(x/2)$ would be rounded to 1 when $x>38$, meaning that $\phi^{-1}\left(\phi(x)\right)=\infty$ for $x>38$ \cite{BrianSPA}. In order to avoid such problems that can arise from limited precision, thresholds on the magnitudes of messages must be applied in simulation studies~\cite{ZhangThesis}.

In order to maintain the performance advantage of SPA decoding over MS decoding, the quantization method has to preserve the self-inverse property of the $\phi(x)$ function and to accurately compute the CN update function in \eqref{eq_SPA_CNphi}.
However, 
it is difficult to have a good approximation of the $\phi(x)$ function with limited resolution, because this requires both fine precision and large range. Efforts have been made to design quantization methods that work effectively with the $\phi(x)$ function. For example, a variable-precision quantization scheme proposed in \cite{Tong} uses larger quantization step size for magnitudes greater than 1, and smaller step size for magnitudes less than 1. An adaptive uniform quantization method proposed in \cite{fixedZhang} uses different quantization step sizes for the outputs of the $\phi(x)$ and the $\phi^{-1}(x)$ function in \eqref{eq_SPA_CNphi}. 
If the output of the $\phi(x)$ function is quantized with finite precision $\epsilon$, inputs greater than $\phi^{-1}(\epsilon)$ can not be distinguished, and $\phi^{-1}(\epsilon)$ is quite small even for extremely fine precision, e.g., $\phi^{-1}(10^{-3})\approx7.6$ and $\phi^{-1}(10^{-6})\approx14.5$.  Hence, the largest supported magnitude during decoding depends on the finest precision of quantization. This means that increasing the quantization range without improving the precision is not beneficial.

In order to avoid dealing with the $\phi(x)$ function, a variety of other CN update rules, most of which are approximations to the SPA, have been proposed. Some of these approximation are based on the following equivalent version of the SPA CN update rule represented by \eqref{eq_SPA_CN} or \eqref{eq_SPA_CNphi},
\begin{equation}
\label{eq_SPA_CNbox}
L_{j\rightarrow i} = \mathop \boxplus \limits_{i'\in N(j)\setminus i}L_{i'\rightarrow j}
\end{equation}
where $\boxplus$ is the pairwise ``box-plus'' operator defined as
\begin{eqnarray}
x\boxplus y &=& \log\left(\frac{1+e^{x+y}}{e^x+e^y}\right)\nonumber\\
&=& \text{sign}(x)\text{sign}(y)\cdot\left\{ \min(|x|,|y|) + s(|x|,|y|) \right\}\label{eq_boxplus_m}\\
&=& \text{sign}(x)\text{sign}(y)\min(|x|,|y|) + s(x,y)\label{eq_boxplus_p}
\end{eqnarray}
with
\begin{equation}
\label{eq_s}
s(x,y)=\log\left(1 + e^{-|x+y|}\right) - \log\left(1 + e^{-|x-y|}\right).
\end{equation}


The proof of equivalence between \eqref{eq_SPA_CN} and \eqref{eq_SPA_CNbox} can be found in \cite{boxplusBP}. We call such an implementation \emph{box-plus} SPA decoding. The formulation above does not have the precision problem that \eqref{eq_SPA_CN} and \eqref{eq_SPA_CNphi} have, and, in fact, in 64-bit double-precision floating-point implementation, the maximum magnitude of a message that can be supported is approximately $1.8\times10^{308}$, which is the largest double-precision value supported by the IEEE 754 standard. Moreover, unlike the $\phi(x)$ function, the function $\log\left(1+e^{-|x|}\right)$ can be well quantized or approximated with piecewise linear functions \cite{BrianSPA,boxplusBP,SPAlinear}.

If the term $s(x,y)$ is omitted when using \eqref{eq_SPA_CNbox} to calculate the CN output in box-plus decoding, the result is the same as that produced by the MS algorithm using~\eqref{eq_MS_CN}. Therefore, box-plus SPA decoding can be viewed as MS decoding with a correction factor. It is known that the magnitude of $s(x,y)$ is bounded above by $\log 2$ (see, for example, \cite[p.~232]{RyanLin}).
In fact, as shown in~\cite{MinSum_vari},\cite{Fossorier}, given the same inputs, a message produced by a CN in SPA decoding has the same sign as the corresponding message in MS decoding, with equal or smaller magnitude. Because of their relevance to the proof of Theorem~\ref{Thm_SPA} below, we summarize these observations relating the CN updates  produced by the SPA and MS decoders in Lemma~\ref{Lemma_MS_vs_SPA}.

\begin{lem}
\label{Lemma_MS_vs_SPA}
Let $L_{MS}$ denote the message from CN $j$ to VN $i$ as computed in (\ref{eq_MS_CN}), and
let $L_{SPA}$ denote the message from CN $j$ to VN $i$ as computed in (\ref{eq_SPA_CN}), (\ref{eq_SPA_CNphi}), and (\ref{eq_SPA_CNbox}). Then
$\text{sign}(L_{SPA}) =\text{sign}(L_{MS})$ and $|L_{SPA}|\leq|L_{MS}|.$
The correction term $s(x,y)$ in (\ref{eq_s}) satisfies
$-\log2 < s(x,y) < \log2$,
and $\text{sign}(s(x,y))=-\text{sign}(xy)$ when $xy\neq 0$.
\end{lem}
\begin{IEEEproof}
See Appendix~\ref{Appendix_Lemma_MS_vs_SPA}.
\end{IEEEproof}
Finally, we note that if the correction term $s(x,y)$ is replaced with a fixed constant, the resulting CN update rule corresponds to that of the OMS decoder in (\ref{eq_OMS_CN}).

As we discussed earlier, no matter how one designs the fixed-point implementation of the original SPA using the $\phi(x)$ function, or even with the floating-point implementation, the function $\left|x-\phi^{-1}\left(\phi(x)\right)\right|$ is unbounded. Even if we saturate both the input and the output of the $\phi(x)$ function, the value of $\left|x-\phi^{-1}\left(\phi(x)\right)\right|$ is still unbounded and linear in $x$. Therefore, the CN output of a practical implementation of \eqref{eq_SPA_CN} or \eqref{eq_SPA_CNphi} can significantly differ from the true computed value. However, since box-plus SPA decoding can be considered as min-sun decoding with a correction factor, the implementation error mainly comes from the computation and quantization of the correction factor, which is a small bounded value, as shown in Lemma~\ref{Lemma_MS_vs_SPA}.
Now, we can extend Theorem~\ref{Thm_MS} to SPA decoding.

\begin{thm}\label{Thm_SPA}
Let $G$ be the Tanner graph of a variable-regular LDPC code that contains a subgraph $S$ induced by a trapping set.
Assume that the channel is either a BSC or an AWGNC, and that the messages from the channel to all VNs outside $S$ are correct.
If $S$ satisfies the $k$-separation assumption for sufficiently large $k$, then the SPA decoder will successfully correct all erroneous VNs in $S$.
\end{thm}
\begin{IEEEproof}
See Appendix~\ref{Appendix_Thm_SPA}.
\end{IEEEproof}

\begin{remk}\label{rm3}
As will be shown in the simulation results, linear scaling of the input LLRs to the SPA decoder will indeed affect the decoding performance, because the correction factor $s(x,y)$ is not linear in either $x$ or $y$.
\end{remk}




For most LDPC codes, the trapping sets typically satisfy the $k$-separation assumption only for small values of $k$. Nevertheless, as described more fully in Section~\ref{sec:nr}, in our 64-bit double-precision floating-point computer simulations of MS decoding and box-plus SPA decoding applied to several LDPC codes traditionally associated with high error floors, we have not observed, in tens of billions of channel realizations of both the BSC and the AWGNC, any decoding failure in which the error patterns correspond to the support of a small trapping set.
Moreover, when we force every VN in a trapping set to be in error and all other VNs to be correct, the floating-point decoders can successfully decode, whereas a decoder implementation that limits the magnitude of messages may not be able to resolve the errors in the trapping set and would then fail to decode to the correct codeword.

We emphasize that the analytical and numerical results in this paper are mainly for variable-regular LDPC codes. Extension of this analysis to variable-irregular LDPC codes does not appear to be straightforward. 

\section{New Quantized Decoders with Low Error Floors}
\label{sec:QuanDec}

As mentioned above, several empirical studies have shown that the range and the precision of quantized messages in iterative LDPC decoders can influence the observed error floor.  Moreover, analytical models used to study the dynamical evolution of messages show that message magnitudes can exhibit exponential growth behavior as a function of the number of decoder iterations.  Likewise, the proofs of the theorems and corollaries in Section~\ref{sec:EF} suggest that iterative decoder performance can be improved by allowing for the exponential growth of message magnitudes.
These results serve as the motivation for a new quantization method that we refer to as $(q+1)$-\emph{bit quasi-uniform} quantization, which we now describe.

Consider first the uniform quantizer with quantization step $\Delta$. For any real number $x$, it is defined by
$$
Q_{\Delta}(x)=sgn(x) \Delta \left\lfloor \frac{|x|}{\Delta}+\frac{1}{2} \right\rfloor.
$$
The outputs of the uniform quantizer are of the form $m\Delta$. The quantization intervals can be visualized by expressing
the quantization rule as
\begin{equation}
\label{eq_uniform}
Q_{\Delta}(x)=\left\{
\begin{array}{ccccc}
m\Delta,& \mbox{\rm if}  & m\Delta -\frac{\Delta}{2} \leq x < m\Delta+\frac{\Delta}{2} & \mbox{\rm for}& m>0 \\
0,      & \mbox{\rm if}  & -\frac{\Delta}{2} < x < \frac{\Delta}{2} &&\\
m\Delta,& \mbox{\rm if}  & m\Delta -\frac{\Delta}{2} < x \leq m\Delta+\frac{\Delta}{2} & \mbox{\rm for}& m<0.\\
\end{array}
\right.
\end{equation}

Now, let $N=2^{q-1}-1$, where $q$ is an integer value $q \geq 1$. The $q$-bit uniform quantizer combines the uniform quantization intervals corresponding to the output values $m\Delta, \; \forall m \geq N$ into a single semi-infinite interval whose elements are quantized to $N\Delta$ and,
similarly, combines the intervals corresponding to the output values $m\Delta, \; \forall m \leq -N$ into a single semi-infinite interval whose elements are quantized to $-N\Delta$.
Denoting the $q$-bit quantizer with step $\Delta$ by $Q_{\Delta, q}(x)$, we have

\begin{equation}
\label{eq_qbituniform}
Q_{\Delta, q}(x)=\left\{
\begin{array}{ccccc}
N\Delta,& \mbox{\rm if}  & N\Delta -\frac{\Delta}{2} \leq x  & &  \\
m\Delta,& \mbox{\rm if}  & m\Delta -\frac{\Delta}{2} \leq x < m\Delta+\frac{\Delta}{2} & \mbox{\rm for}& N>m>0 \\
0,      & \mbox{\rm if}  & -\frac{\Delta}{2} < x < \frac{\Delta}{2} && \\
m\Delta,& \mbox{\rm if}  & m\Delta -\frac{\Delta}{2} < x \leq m\Delta+\frac{\Delta}{2} & \mbox{\rm for}& 0>m>-N.\\
-N\Delta,& \mbox{\rm if}  &  x \leq  -N\Delta + \frac{\Delta}{2}& &  \\
\end{array}
\right.
\end{equation}

The number of intervals is $2N+1 = 2^q-1$, and the quantizer output levels $m\Delta, \; -N \leq m \leq N$, can be denoted by the signed
$q$-bit binary representation of $m$, that is,  $[m_0, m_1, \ldots, m_{q-1}]$, where the last $q-1$ bits are the binary representation of $|m|$,
and $m_{0}$ is the sign bit with value 0 (resp. 1) when $m$ is positive (resp. negative). Note that the output level 0 has two such
binary representations; one of them can be selected using any preferred convention.

One approach to expanding the range of quantized messages is to increase the step size $\Delta$, without changing the resolution $q$.
This approach, however sacrifices the precision of the quantization. Alternatively, one could maintain the value of $\Delta$ and increase
$q$ to resolve larger magnitudes. This would increase implementation complexity when incorporated into the decoding hardware.

In the context of our application, the $(q+1)$-\emph{bit quasi-uniform} quantizer represents a compromise between these conflicting objectives of retaining fine precision, allowing large dynamic range, and controlling implementation complexity as messages grow exponentially in the number of decoder iterations. The definition of the quantizer involves another parameter $d>1$, which we refer to as the \emph{growth rate} parameter.
Roughly speaking, the underlying idea behind the quantizer is as follows. For input values in the interval $(-dN\Delta, dN\Delta)$,we use $q$-bit uniform quantization with step size $\Delta$. The intervals corresponding to quantized values $m\Delta, \; -(N-1) \leq m \leq (N-1)$ are exactly like those of the $q$-bit uniform quantizer. For values $N\Delta$and $-N\Delta$, the semi-infinite intervals are shortened to have length $dN\Delta - N\Delta + \frac{\Delta}{2}.$  For input values with magnitude larger than $dN\Delta$, the quantizer outputs can take an additional $N+1=2^{q-1}$ values of the form $d^r N\Delta, \; 1\leq r \leq N+1$, with corresponding intervals that increase exponentially in length with growth rate $d$.
More precisely, the  $(q+1)$-\emph{bit quasi-uniform} quantizer, denoted by $Q^*_{\Delta,q}(x)$ is defined as follows.

\begin{equation}
\label{eq_quasiuniform}
Q^*_{\Delta, q}(x)=\left\{
\begin{array}{ccccc}
d^{N+1} N\Delta,& \mbox{\rm if}  & d^{N+1}N\Delta \leq x  & &  \\
d^r N\Delta,& \mbox{\rm if}  & d^r N\Delta \leq x < d^{r+1}  N\Delta,& \mbox{\rm for}& N \geq r \geq 1  \\
Q_{\Delta, q}(x), & \mbox{\rm if}  & - d N\Delta < x  < d N\Delta& &\\
-d^r N\Delta,& \mbox{\rm if}  & -d^{r+1}N\Delta  < x \leq -d^r N\Delta,& \mbox{\rm for}& 1 \leq r \leq N  \\
-d^{N+1} N\Delta,& \mbox{\rm if}  & x \leq -d^{N+1}N\Delta   & &  \\
\end{array}
\right.
\end{equation}a

From Definition~\eqref{eq_quasiuniform}, we see that the quantization levels can be represented with only $q+1$ bits.
The levels $m\Delta, \; -N \leq m \leq N$ are represented by $[m_0, m_1, \ldots, m_{q-1}, m_q]$, where $[m_0, m_1, \ldots, m_{q-1}]$
is the signed binary representation of the integer $m$ and the final \emph{indicator bit},
$m_q$  is set to zero, i.e., $m_q=0$, to reflect the fact that the $q$-bit uniform quantizer has been applied.
The $2^{q}$ quantized levels $d^r N\Delta, \; 1 \leq r \leq N+1$ are denoted by $[r_0, r_1, \ldots, r_{q-1},r_q ]$, where
$[r_0, r_1, \ldots, r_{q-1}]$ is the signed binary representation of $r-1$, and the indicator bit $r_q$ is set to 1, i.e., $r_q=1$, to indicate that
non-uniform quantization has been used. Similarly, we denote the $2^{q}$ quantized levels $- d^r N\Delta, \; 1 \leq r \leq N+1$
by $[r_0, r_1, \ldots, r_{q-1}, r_q]$, where  $[r_0, r_1, \ldots, r_{q-1}]$ is the signed binary representation of $-(r-1)$, and the indicator bit $r_q$ is again set to 1, i.e., $r_q=1$. It is sometimes convenient to represent these quantization levels in the form $(y,b)$, where $y$ is the decimal integer representation of the signed binary $q$-tuple $[m_0, m_1, \ldots, m_{q-1}]$ or $[r_0, r_1, \ldots, r_{q-1}]$,  and $b$ is the indicator bit
$m_q$ or $r_q$.

Table~\ref{tab_31bit} shows an example of (3+1)-bit quasi-uniform quantization with $\Delta=1$, $q=3$, and $d=3$. Here $N=3$. The operation of the quantizer is shown only for non-negative real inputs. The operation on negative reals can be obtained by odd symmetry. The first bit is the sign bit, and the last bit is the indicator bit. The quantizer behaves just like the 3-bit uniform quantizer in the interval $[0, 9)$. When $x \geq 9$,
the quantizer uses intervals of exponentially increasing length, with input $x$ quantized to the smallest value in the interval in which $x$ falls.
For example, all values within the quantization interval $[27,81)$ are quantized to 27. The decimal values are used in the VN and CN update computations, and then the corresponding quantized binary messages are passed between VNs and CNs.

\begin{table}
\centering
\begin{minipage}[b]{0.45\linewidth}
\caption{(3+1)-bit quasi-uniform quantization with $\Delta=1$ and $d=3$.}
\begin{tabular}{|c|c|c|}
\hline
Input & Quantized value & Binary \\
range & (decimal) & form\\
\hline
[0,0.5] & 0 & 0000 \\ \hline
(0.5,1.5] & 1 & 0010 \\ \hline
(1.5,2.5] & 2 & 0100 \\ \hline
(2.5,9) & 3 & 0110 \\ \hline
[9,27) & 9 & 0001  \\ \hline
[27,81) & 27 & 0011 \\ \hline
[81,243) & 81 & 0101 \\ \hline
[243, $\infty$) & 243 & 0111\\ 
\hline
\end{tabular}
\label{tab_31bit}
\end{minipage}
\centering
\hspace{0.5cm}
\begin{minipage}[b]{0.45\linewidth}
\caption{4-bit quasi-uniform quantization with $\Delta=1$, $d=3$, and $N_u=5$.}
\begin{tabular}{|c|c|c|}
\hline
Input & Quantized value & Binary \\
range & (decimal) & form\\
\hline
(0,0.5] & 0 & 0000 \\ \hline
(0.5,1.5] & 1 & 0001 \\ \hline
(1.5,2.5] & 2 &  0010 \\ \hline
(2.5,3.5] & 3 & 0011  \\ \hline 
(3.5,12) & 4 & 0100 \\ \hline
[12,36) & 12 & 0101  \\ \hline 
[36,108) & 36 & 0110  \\ \hline
[108,$\infty$) & 108 & 0111  \\ 
\hline
\end{tabular}
\label{tab_4bit}
\end{minipage}
\end{table}


We can further extend the idea of ($q$+1)-bit quasi-uniform quantization, as follows. The $(q+1)$-bit quasi-uniform quantizer uses $q+1$ bits in total to represent  $2N+1=2^q$ different output magnitudes, or $2^{q+1}-1$ quantization intervals if signs are taken into account. As described in \eqref{eq_quasiuniform} and illustrated in Table~\ref{tab_31bit}, $N+1=2^{q-1}$ output magnitudes, including 0, are allocated to the uniform quantization domain and the remaining $N+1=2^{q-1}$ magnitudes correspond to exponentially growing quantization interval lengths.
The generalized (symmetric) $(q+1)$-bit quasi-uniform quantizer represents the same number of magnitudes, but it can assign any number, say $N_u$,  to the uniform quantization range and the remaining $2^q-N_u$ magnitudes to the exponential quantization range. With a quantization rule similar to \eqref{eq_quasiuniform}, the quantized values of the general $q+1$-bit quasi-uniform quantization are $m\Delta$ for $-N_u< m < N_u$; $d^{r-N_u+1}N_u\Delta$ for $r\geq N_u$, and $-d^{r-N_u+1}N_u\Delta$ for $r\leq-N_u$.
Table~\ref{tab_4bit} shows an example of a general 4-bit quasi-uniform quantization with $N_u=5$, $\Delta=1$, and $d=3$.
The uniform quantization range in this example is from $-4$ to 4 with uniform step size 1, and the exponential range is above 4 or below $-4$ with exponential step sizes $4\cdot(3^r-3^{r-1})$ for $1\leq r\leq3$.

The motivation for the proposed quasi-uniform quantization method was the analysis of message-passing decoder behavior on trapping sets that satisfy the $k$-separation assumption for large $k$. Although this property is generally not satisfied by trapping sets in practical LDPC codes, the simulation results in the next section demonstrate that, for a variety of LDPC codes that were examined, this quantization approach can substantially lower error floors when used with standard MS-based and SPA-based decoders.

\section{Numerical Results}\label{sec:nr}
In this section we compare the error-rate performance obtained with the proposed quasi-uniform quantization method to that obtained using uniform quantization.
We consider four know LPDC codes covering a range of rates and lengths:  a rate-$\frac{3}{10}$, (640,192) quasi-cyclic (QC) LDPC code \cite{han_ryan}; the rate-$\frac{1}{2}$, (2640,1320) Margulis LDPC code \cite{nearcodeword}; the rate-$\frac{4}{5}$, (1280, 1024) AR4JA LDPC code \cite{Hamkins}; and MacKay's (4095,3358) regular LDPC code (the 4095.737.3.101 code in \cite{MackayCode}) with rate approximately 0.82.  Results are shown for various combinations of the BSC and AWGN channels using the MS, OMS, AMS, SPA, and approximated-SPA decoders.

All of the frame error rate (FER) curves are based on Monte Carlo simulations that generated at least 200 error frames for each plotted error rate, and the maximum number of decoding iterations was set to 200, unless otherwise indicated.

\subsection{Dynamical Range of Message Magnitudes}\label{sec:dr}

We first present some empirical data in support of the contention that some benefit may come from allowing message magnitudes to grow during iterative decoding.

\begin{figure}[!t]
\centerline
{\subfigure[MS decoder on the (640,192) QC-LDPC code over BSC of $p=0.03$ and $|LLR|=1$.]{\includegraphics[width=0.45\linewidth]{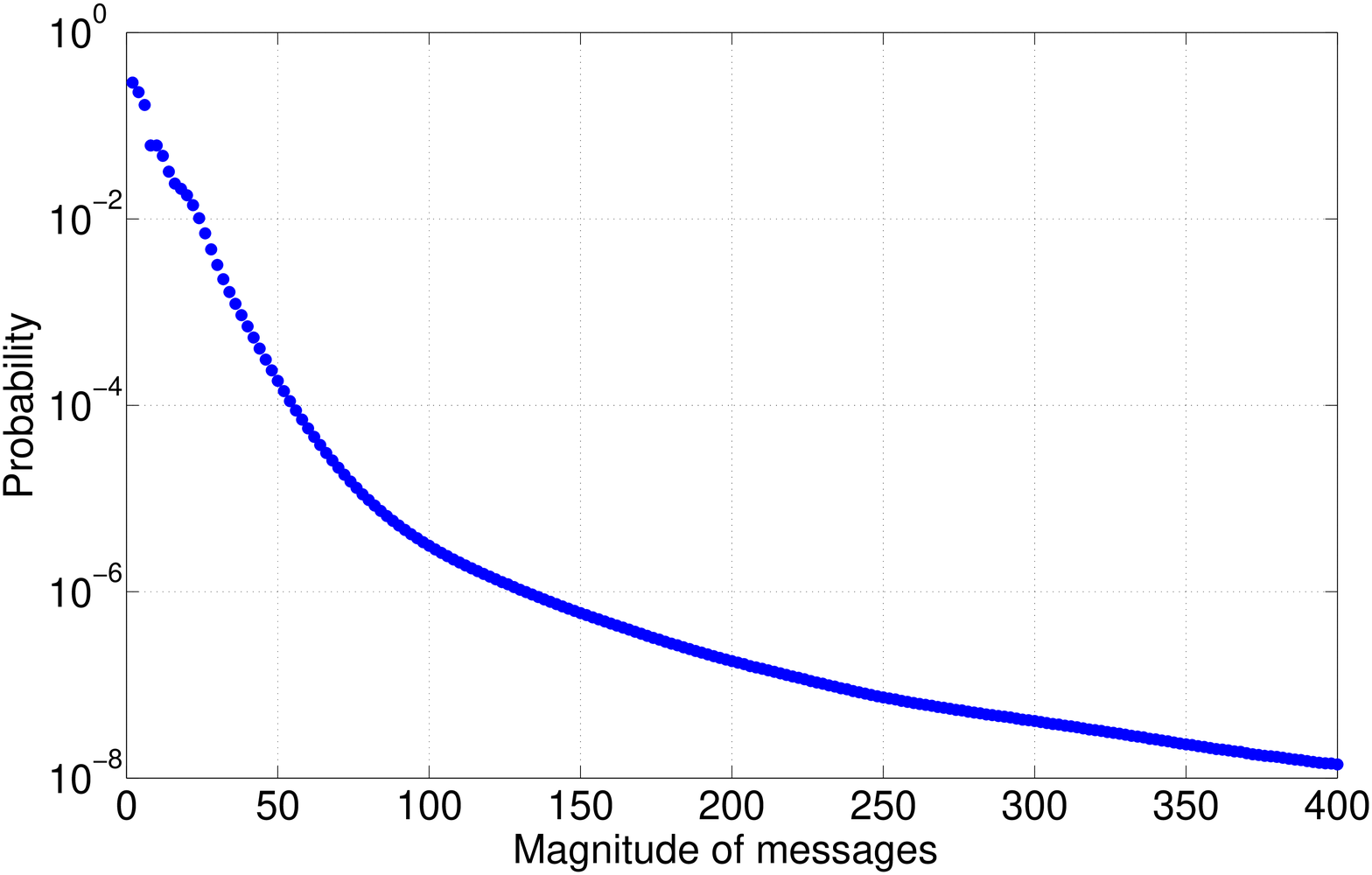}
\label{subfig_pdf640}}
\hfil
\subfigure[SPA decoder on the Margulis code of length 2640 over AWGNC of $E_b/N_0=2.25$ dB.]{\includegraphics[width=0.45\linewidth]{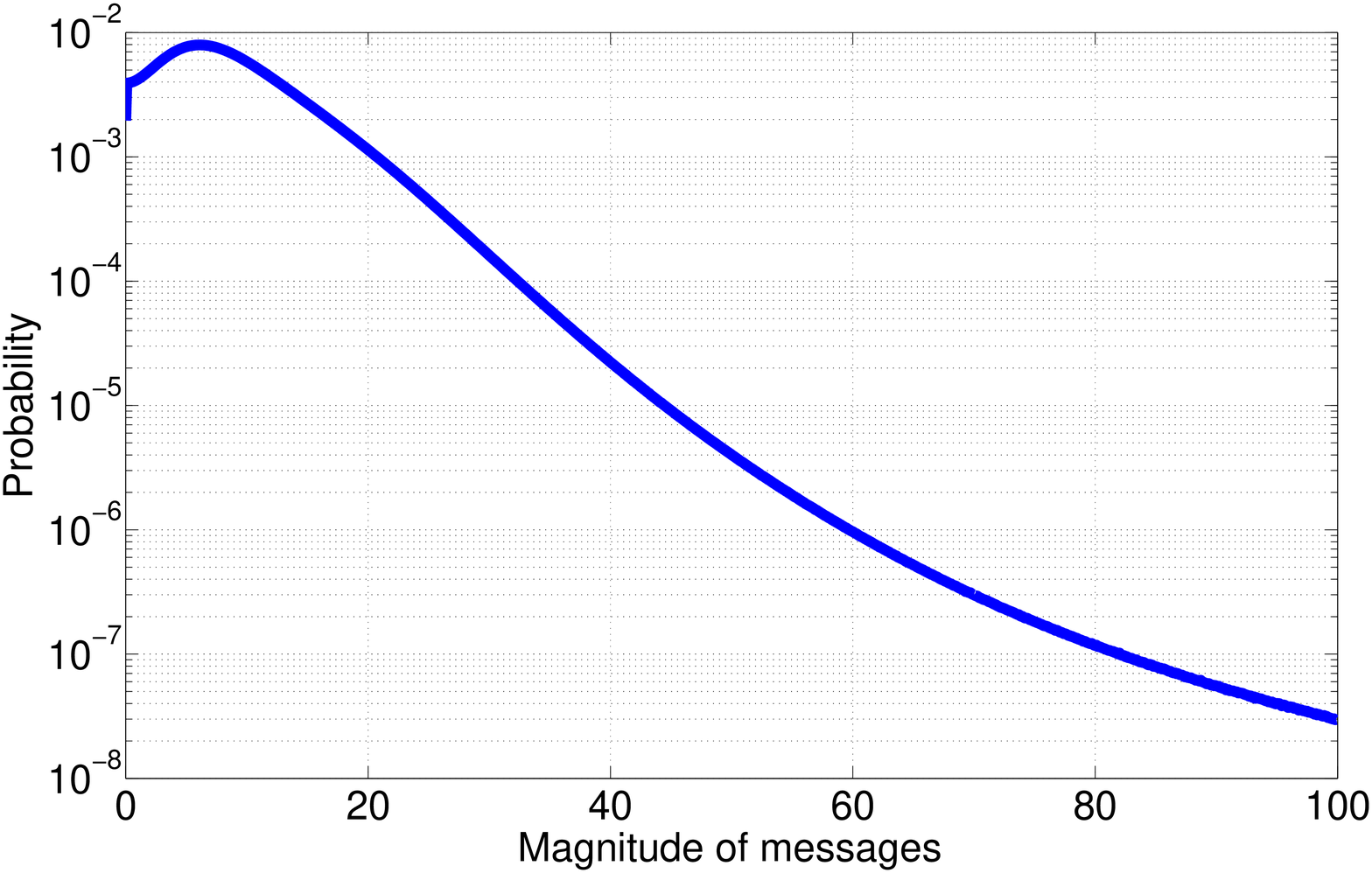}
\label{subfig_pdf2640}}}
\caption{Probability density function of magnitude of messages.}
\label{fig_pdf}
\end{figure}

Fig.~\ref{fig_pdf} shows the empirical probability density functions (pdf) of the message magnitudes observed during decoding simulations for two LDPC codes.
Fig.~\ref{subfig_pdf640} shows the pdf for the MS decoder applied to the (640,192) QC-LDPC code on the BSC with $p=0.03$, where the magnitude of all input LLRs is scaled to 1.
Fig.~\ref{subfig_pdf2640} shows the pdf of the SPA decoder applied to the Margulis code on the AWGNC with $E_b/N_0=2.25$ dB.
The data used to create these figures were obtained using floating-point decoder implementations and more than 10 million channel output symbols. The messages passed on all edges during all decoding iterations were collected to generate the pdfs.
In the simulations, the iterative MP decoders stopped when a codeword was found or when the maximum number of iterations (200) was reached. The figures confirm that a substantial fraction of messages had ``large'' magnitudes. Moreover, upon further examination of the simulation data, we found that such ``strong'' messages, in general, helped to successfully decode the received symbols, as suggested by the idealized theoretical analysis in Section \ref{sec:EF}.

\begin{figure}
\begin{minipage}[b]{0.49\linewidth}
\includegraphics[width=\textwidth]{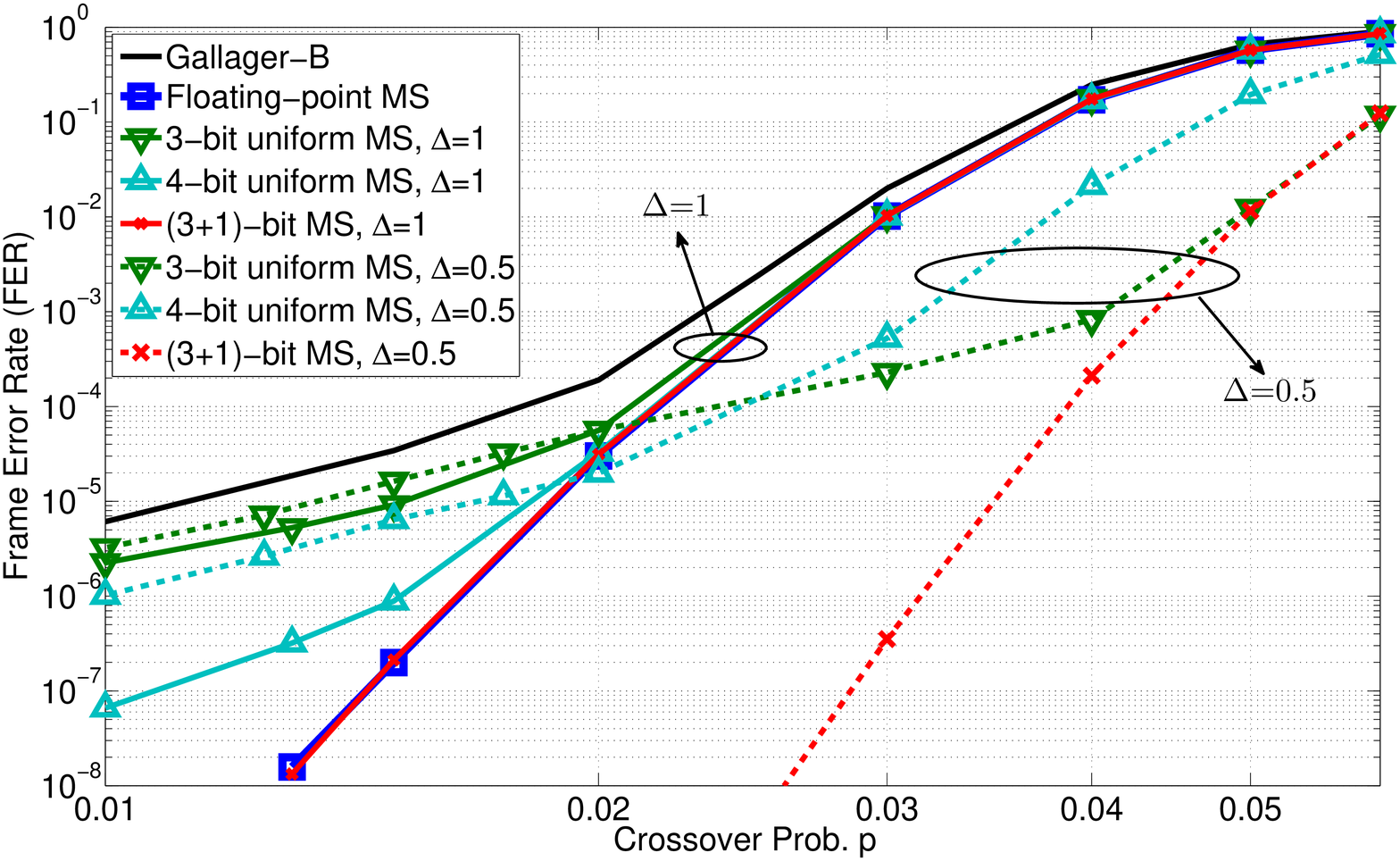}
\centering
\caption{FER results of min-sum (MS) decoder on the (640,192) QC-LDPC code on BSC, where $\Delta=1$ or $0.5$, and $d=3$.}\label{fig_bsc640ms}
\end{minipage}
\hspace{0.1cm}
\begin{minipage}[b]{0.49\linewidth}
\includegraphics[width=\textwidth]{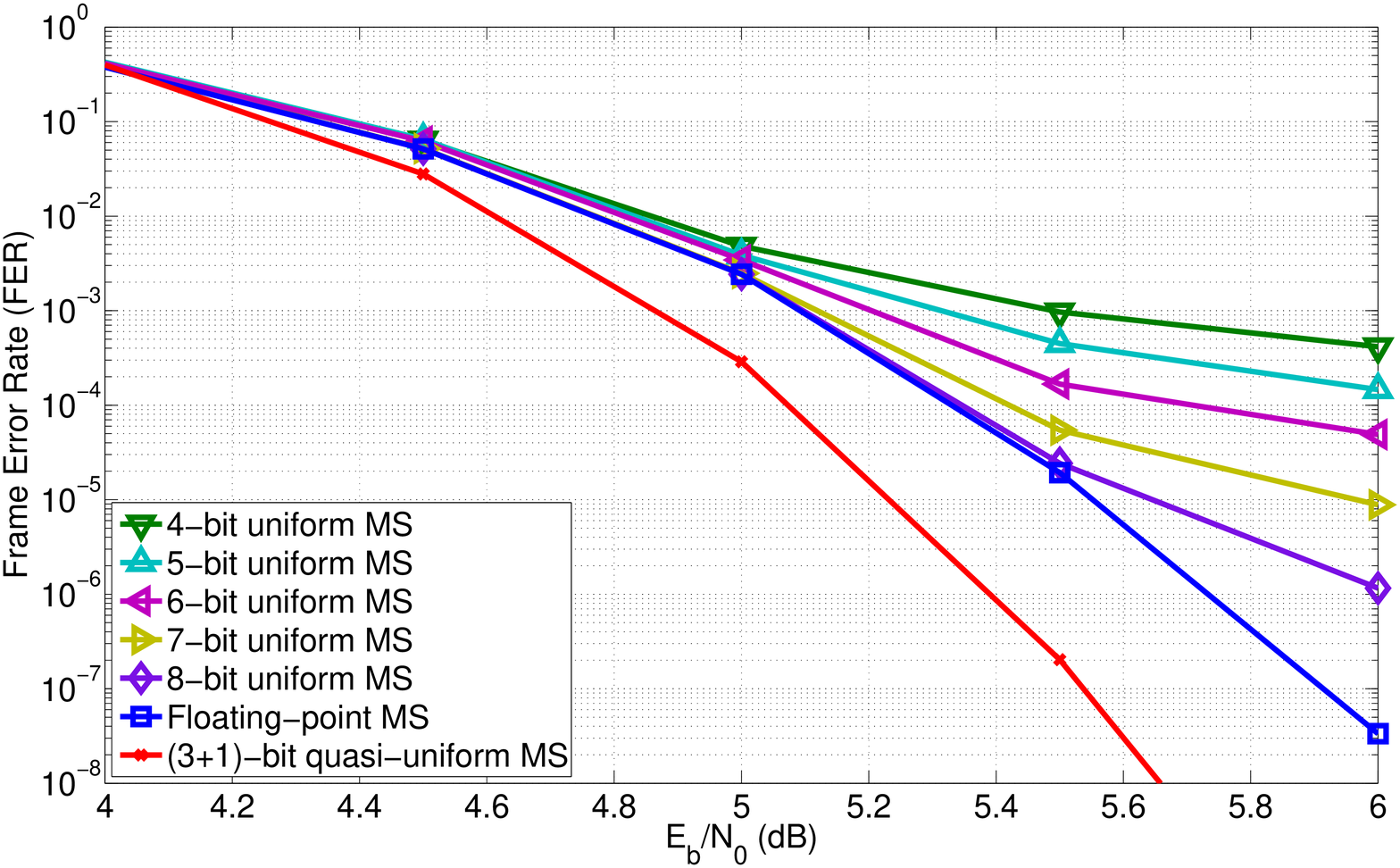}
\centering
\caption{FER results of min-sum (MS) decoder on the (640,192) QC-LDPC code on AWGNC, where $\Delta=0.5$ and $d=3$.}\label{fig_awgn640ms}
\end{minipage}
\end{figure}

\subsection{Simulation Results for Min-Sum Decoding and Variants}\label{sec:sim_ms}

Figs.~\ref{fig_bsc640ms} and \ref{fig_awgn640ms} show simulation results for the (640,192) QC-LDPC code using various types of quantized MS decoders and floating-point MS decoders, extending some of the results presented in \cite{ISIT_MS}.
For the BSC, we scaled the magnitudes of decoder input messages from the channel to 1 since, for linear decoders such as Gallager-B and MS, the scaling of channel input messages does not affect the decoding performance.
The uniform quantization step size $\Delta$ is set to 1 or 0.5. So, for example, when $\Delta=1$, the 3-bit uniform quantizer produces values\ $\{\pm3,\pm2,\pm1,0\}$, and the (3+1)-bit quasi-uniform quantizer with $d=3$ yields the values described in Table~\ref{tab_31bit}. In the simulation, the parameter $d$ was heuristically chosen by testing different values. When $q$ is large, a small $d$ proved to be enough to represent a large range of message magnitudes.

In Fig.~\ref{fig_bsc640ms}, we see that the slope of the error floor resulting from uniform quantization with either step size, $\Delta=1$ or $\Delta=0.5$, is similar to that of the Gallager-B decoder error floor. This is because, when most messages have the same magnitude, MS decoding essentially degenerates to Gallager-B decoding, which relies solely upon the signs of messages.

Comparing uniform quantizers with the same number of bits but different step sizes, we see that smaller step size produces better performance in the waterfall region but a higher error floor. This observation can be explained by the saturation level of these quantizers. For example, 3-bit and 4-bit uniform quantizers with step size $\Delta=1$ saturate at magnitudes 3 and 7, respectively, whereas with step size $\Delta=0.5$, they saturate at magnitudes 1.5 and 3.5, respectively. The stronger messages, i.e., the messages with larger magnitudes, can be helpful or harmful to the decoding process, depending on whether they are correct or not. The correct ones can help overcome the incorrectly received bits, but the incorrect ones tend to  negatively influence the recovery of correctly received bits. In the error-floor region, when channel conditions are good, very few bits are received incorrectly, and as suggested by the proofs of Theorems \ref{Thm_MS} and \ref{Thm_SPA}, large saturation levels allow messages corresponding to correct bits to grow sufficiently to overcome the ``incorrect'' messages in trapping sets. This behavior is evident in Fig.~\ref{fig_bsc640ms}, where the error floors produced by the different uniform quantizers monotonically decrease as the saturation levels increase.

On the other hand, in the waterfall region where many bits are received incorrectly, reducing the saturation level may limit the propagation of strong incorrect messages. Moreover, in this specific case, quantization with the smaller step size $\Delta=0.5$ may be expected to improve performance relative to that achieved with the larger step size $\Delta=1$ or with a floating-point MS decoder implementation.
The reasoning is that, since the magnitudes of input LLRs to the MS decoder from the BSC are scaled to 1, the low saturation level and the possible appearance of non-integral saturated messages may reduce the possibility of the messages at a VN summing to zero. Because having VNs summing to zero could result in oscillatory behavior in the decoder and failure to decode correctly, this could explain why in Fig.~\ref{fig_bsc640ms} the MS decoder using (3+1)-bit quasi-uniform quantization and step size $\Delta=0.5$ yields better performance than the floating-point decoder.

Fig.~\ref{fig_awgn640ms} shows the performance of MS decoding of the (640,192) QC-LDPC code on the AWGNC channel. Here the $(3+1)$-bit quasi-uniform quantizer yields  substantial reduction of the error floor in comparison not only to 8-bit uniform quantization but also to the floating-point results. This is consistent with, and more impressive than, the results shown in \cite{ISIT_MS} for the Margulis code, where $(5+1)$-bit quasi-uniform quantization surpassed 6-bit uniform quantization and paralleled floating-point results.
Heuristic reasoning along the lines used above suggests that codes with higher variable-node degree would benefit even more from the quasi-uniform quantization.
However, it is important to point out that the gains can be code-dependent, so further performance studies are needed to confirm this.

\begin{figure}
\begin{minipage}[b]{0.49\linewidth}
\includegraphics[width=\textwidth]{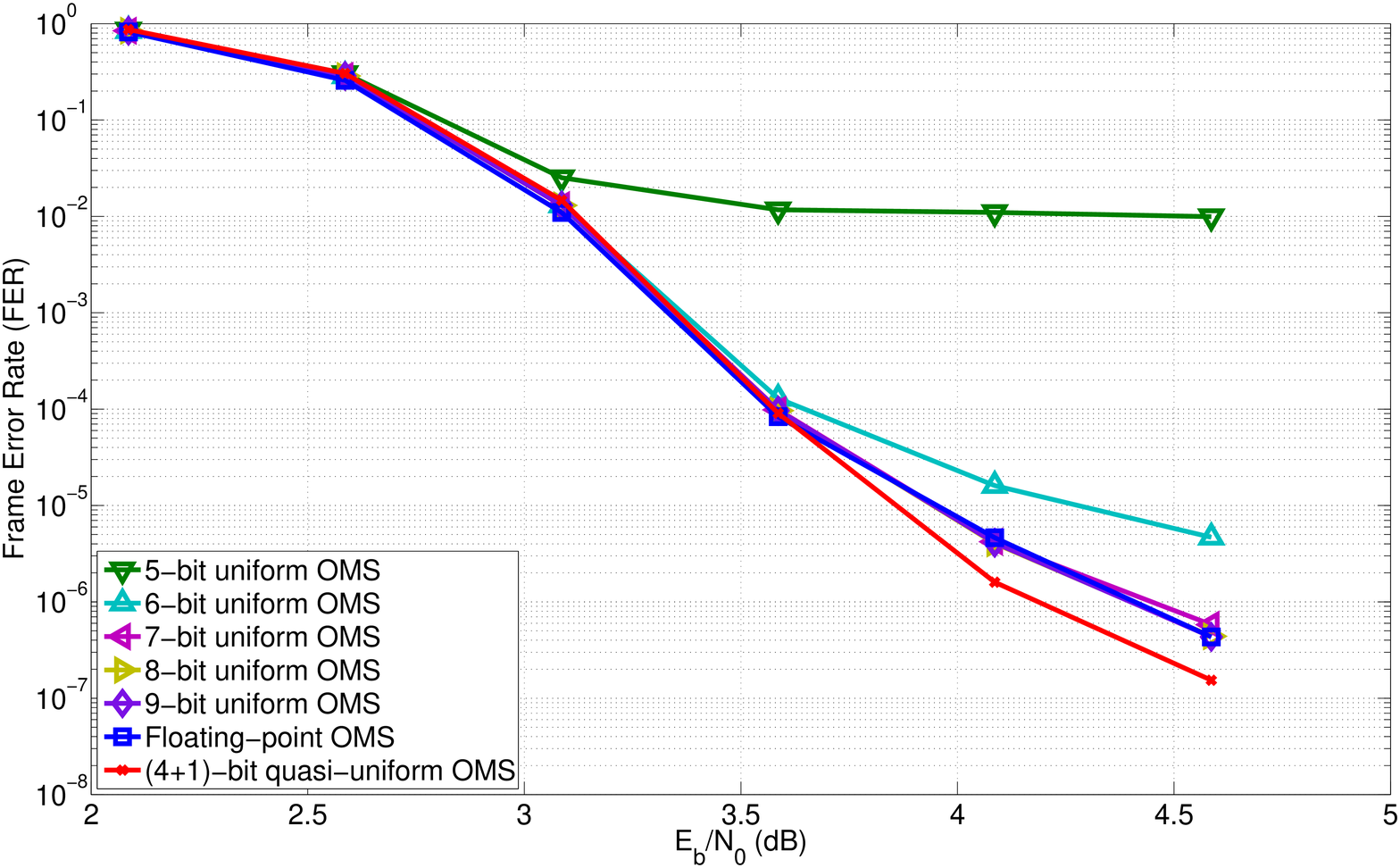}
\centering
\caption{FER results of OMS decoder on the AR4JA LDPC code of $k=1024$ and $r=0.8$ on AWGNC, where $\Delta=0.5$, $d=1.5$, and offset factor $\beta=0.5$,.}\label{fig_awgnar4jaoms}
\end{minipage}
\hspace{0.1cm}
\begin{minipage}[b]{0.49\linewidth}
\includegraphics[width=\textwidth]{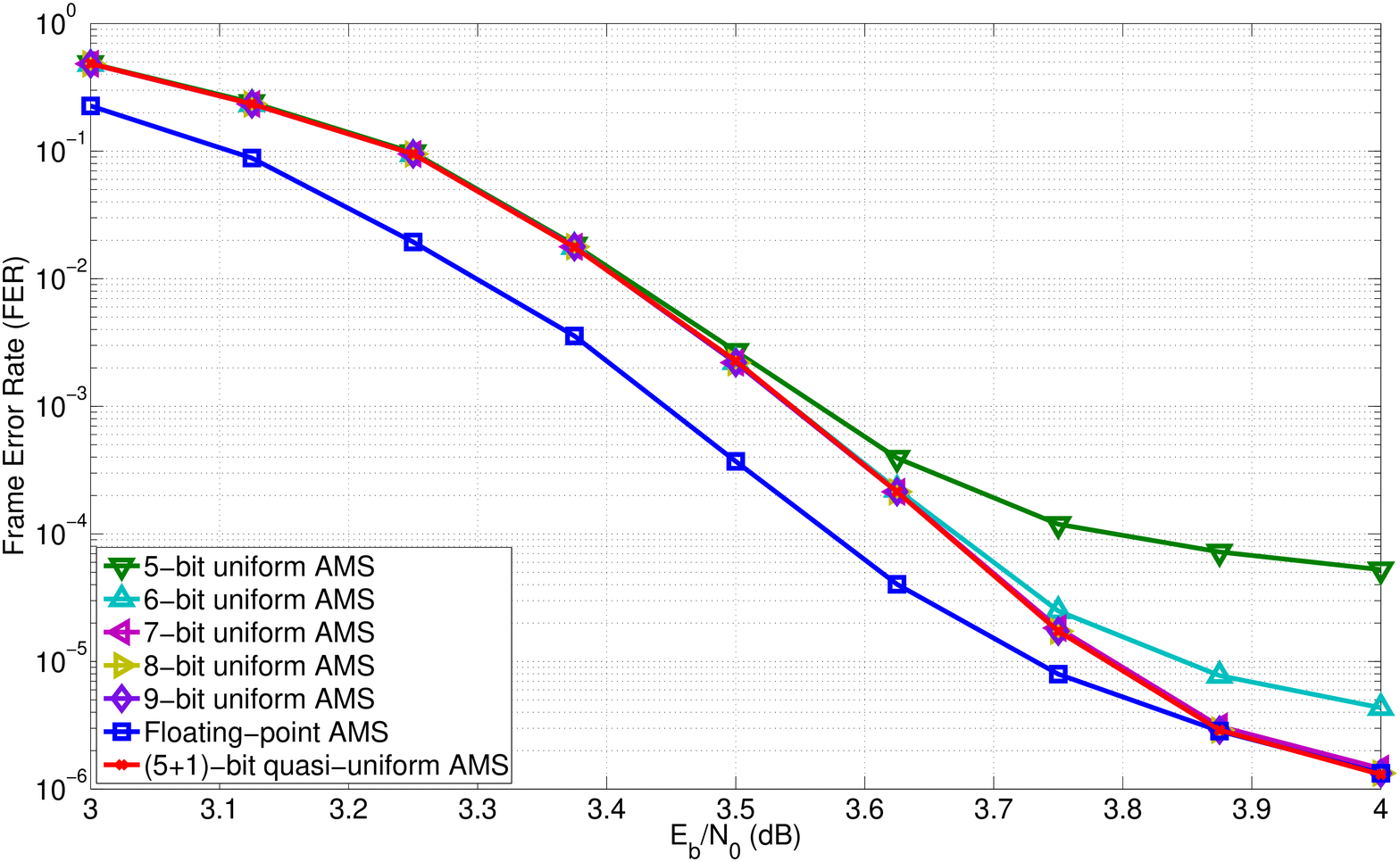}
\centering
\caption{FER results of AMS decoder on the (4095,3358) LDPC code  on AWGNC, where $\Delta=0.5$, $d=1.3$, and attenuation factor $\alpha=0.7$.}\label{fig_awgn4095ams}
\end{minipage}
\end{figure}

Quasi-uniform quantization can be directly applied to modified MS decoders, such as AMS and OMS, with the possibility of significant reduction in the error floor. This was illustrated in \cite{ISIT_MS} for the (640,192) QC-LDPC code with AMS decoding on the BSC and with OMS decoding on the AWGNC. In case of AMS decoding, $(3+1)$-bit quasi-uniform quantization dramatically reduced the error floor relative to 4-bit uniform quantization, achieving the performance of the unsaturated AMS decoder. For OMS decoding with $(4+1)$-bit quantization, the comparisons to 5-bit uniform quantization and unsaturated decoding were analogous.

Here we consider the performance of AMS and OMS decoding on longer codes with higher rates, specifically, the rate-0.8, (4095,3358) regular code and the rate-0.8, (1280, 1024), irregular AR4JA code.
Fig.~\ref{fig_awgnar4jaoms} compares the quasi-uniform quantization method with uniform quantization in OMS decoding. The performance of the floating-point OMS decoder is also shown. With uniform quantization ranging from 5 bits to 9 bits, we can see that 8 bits suffice to closely approach the error-rate performance of floating-point OMS, whereas the (4+1)-bit quasi-uniform quantization actually surpasses floating-point decoder.
Fig.~\ref{fig_awgn4095ams} shows a similar comparison for AMS decoding of MacKay's (4095,3358) LDPC code. The attenuation factor $\alpha$ was set to the value 0.7, which was found empirically to give the best error floor performance among integer multiples of 0.1 in the range [0.5, 0.9]. After normalization by this factor in every CN update, we found that the quantized value lost precision due to the coarse step size $\Delta=0.5$. As a consequence, the floating-point AMS decoder had better performance than any of the quantized decoders, most noticeably in the waterfall region. Uniform quantization with 7 or more bits appears to eventually achieve floating point performance at FER below $3 \times 10^{-6}$, as does $(5+1)$-bit quasi-uniform quantization.

\subsection{Simulation Results for Sum-Product Algorithm Decoding}\label{sec:sim_spa}

We now consider the application of quasi-uniform quantization to SPA decoding. In our simulations of quantized SPA decoding, the input LLRs and the messages passed between CNs and VNs are quantized values. For convenience, the CN updates are carried out with floating-point arithmetic using the box-plus update rule in \eqref{eq_SPA_CNbox}; the resulting message is then quantized appropriately.


\begin{figure}
\begin{minipage}[b]{0.49\linewidth}
\includegraphics[width=\textwidth]{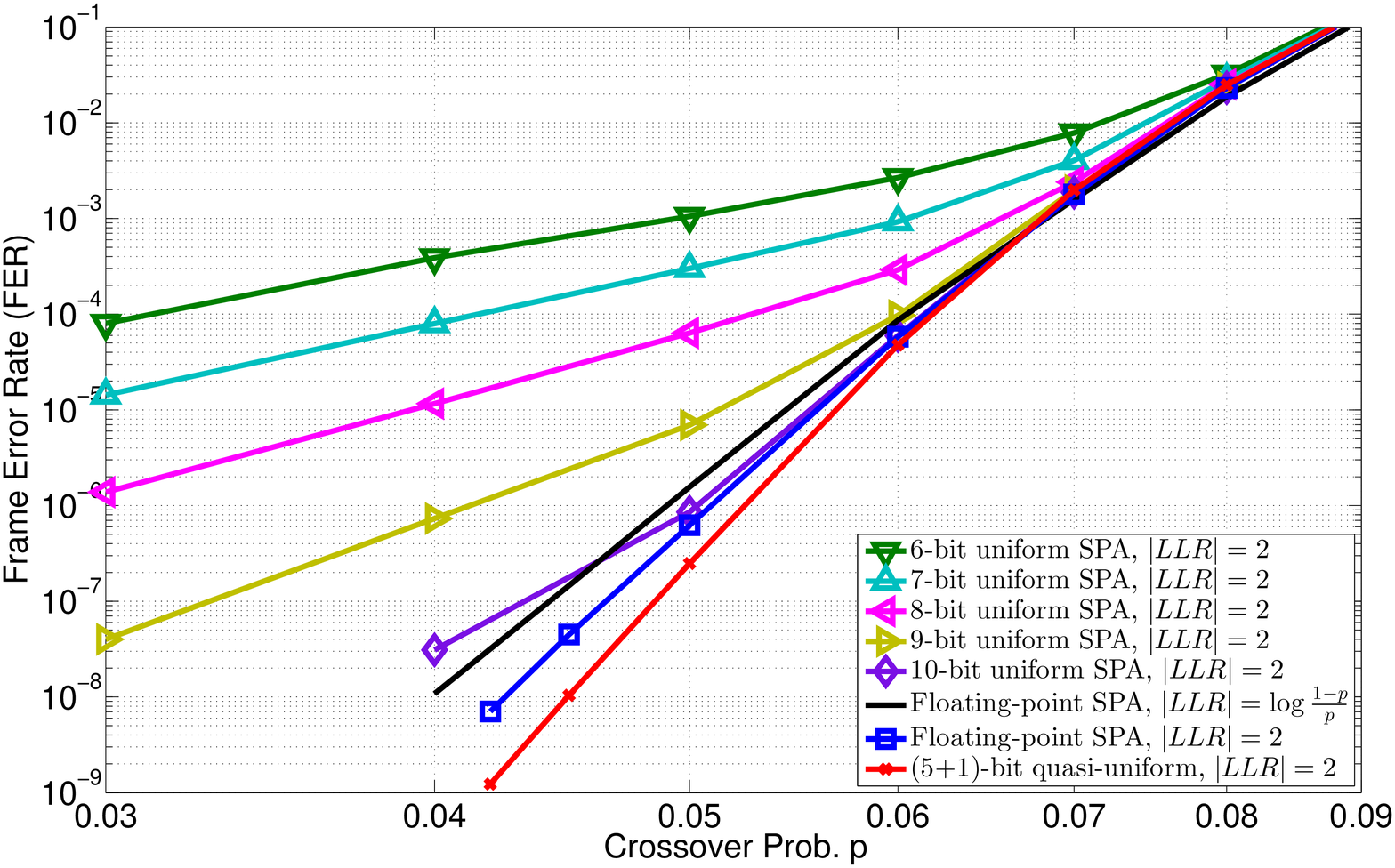}
\centering
\caption{FER results of SPA decoder on the (640,192) QC-LDPC code on BSC, where $\Delta=0.25$, and $d=1.3$.}\label{fig_bsc640spa}
\end{minipage}
\hspace{0.1cm}
\begin{minipage}[b]{0.49\linewidth}
\includegraphics[width=\textwidth]{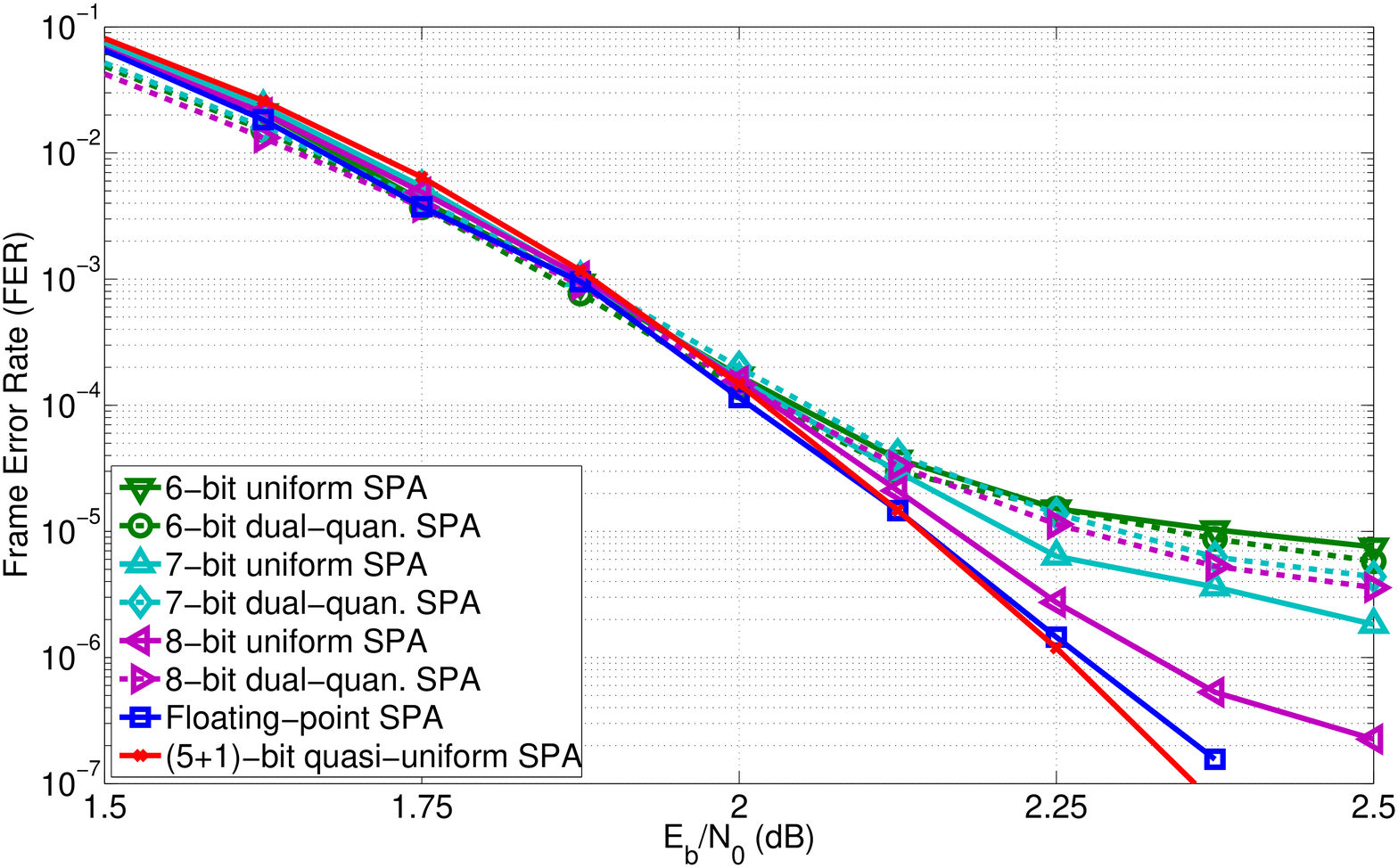}
\centering
\caption{FER results of approx.-SPA decoder on the (2640,1320) Margulis code on AWGNC, where $\Delta=0.25$ and $d=1.3$.}\label{fig_awgn2640spa}
\end{minipage}
\end{figure}

In~\cite{SPCOM_SPA}, we illustrated the performance of quasi-uniform quantization with SPA decoding of the (640,192) QC-LDPC code on the BSC. We saw that with LLR magnitudes scaled to 2, the (6+1)-bit quasi-uniform quantizer with  step size $\Delta=0.25$ and $d=1.5$ performs significantly better than 7-bit uniform quantization with the same step size. Its performance is comparable to that of the floating-point SPA decoder, which is superior to floating-point SPA decoding with exact LLR magnitudes
$\log \frac{1-p}{p}$ when the channel error probability $p$ is small.

Here we consider the same code and channel, with step size again set to $\Delta=0.25$, but with quantization value scale factor reduced to $d=1.3$. With LLR magnitudes scaled to 2, we simulated 6-bit through 10-bit uniform quantization, (5+1)-bit quasi-uniform quantization, and floating-point SPA decoding with LLR magnitudes scaled to 2 as well as with exact LLR magnitudes.

The simulation results, shown in Fig.~\ref{fig_bsc640spa}, indicate that the $(5+1)$-bit quasi-uniform quantizer provides the best performance for $p<0.06$. Comparing to the results in~\cite{SPCOM_SPA}, the performance of the $(5+1)$-bit quantizer with $d=1.3$ is only slightly worse than that of the $(6+1)$-bit quantizer with $d=1.5$.  

We note that the selection of the input LLR magnitude, here set to 2, is heuristic and code-dependent. The value 2 was found empirically to give much better performance than, for example, the value 1, but does not necessarily represent the optimal LLR magnitude scaling.

Results for SPA decoding of the (640,192) QC-LDPC code on the AWGN channel were also presented in~\cite{SPCOM_SPA}. The $(6+1)$-bit quasi-uniform quantizer with $\Delta=0.25$  and $d=1.5$ was found to significantly improve upon 7-bit uniform quantization and match the performance of the floating-point box-plus SPA decoder.


In~\cite{SPCOM_SPA}, we found similar relative performance for the Margulis code on the AWGNC. The $(6+1)$-bit quasi-uniform quantizer outperformed  7-bit uniform quantization, with step size parameters $\Delta=0.25$ and $d=1.2$, and its performance  equaled that of the  ``approximated box-plus SPA'' decoder.  The latter made use of a two-piece linear approximation for $\ln(1+e^{-|x|})$, taken from~\cite{SPAlinear}, in computing the correction factor $s(x,y)$ for box-plus SPA decoding in \eqref{eq_s}, namely,
\begin{equation}
\label{eq_approx}
\ln\left(1+e^{-|x|}\right)=\left\{ \begin{gathered}
  0.6-0.24|x|$,\quad if $|x|<2.5\hfill\\
  0$,\qquad\qquad\qquad otherwise.$\hfill\\
\end{gathered}  \right.
\end{equation}
\noindent
The approximated decoder ran about five times faster than the floating-point SPA decoder, with a performance penalty of less than 0.02~dB in the waterfall region.

In Fig.~\ref{fig_awgn2640spa} we show further results for the Margulis code on the AWGNC.
The plot shows the FER results for $(5+1)$-bit quasi-uniform quantization, as well as  6-, 7-, and 8-bit uniform quantizers, with quantization parameters set to  $\Delta=0.25$ and $d=1.3$.
We also evaluated the dual quantization SPA decoding proposed by Zhang \emph{et al.} \cite{fixedZhang}, where the $\phi(x)$ function is quantized into a mapping table, denoted as $\bar\phi(x)$. Following the notation in \cite{fixedZhang}, we considered dual quantization with parameters Q4.2/1.5, Q5.2/1.6, and Q6.2/1.7 for 6-bit, 7-bit, and 8-bit quantizers, respectively.
The Q$m.f$ quantizer uses uniform quantization to represent a signed fixed-point number with $m$ bits to the left of the radix point for the integer part and $f$ bits to the right of the radix point for the fractional part. For example, a Q4.2 quantizer has uniform quantization step size of 0.25 and a range $[-7.75, 7.75]$.
Hence, all the quantization methods compared here have the same uniform step size of $\Delta=0.25$ when quantizing the input LLRs.

We know that the saturation level $\bar\phi(0)$ is limited by the quantization step size, because it is desirable to have $\bar\phi(0)<x$ for all $x$ satisfying $\bar\phi(x)=0$. In other words, in the dual quantization scheme, the saturation level has to match the resolution of the quantizer; otherwise the error-rate performance in both the waterfall region and the error-floor region will be significantly degraded. Based on error-rate simulations using a range of saturation levels for dual quantization methods, we chose the saturation level for $\bar\phi(x)=0$ to be 5.5, 7, and 8 for the 6-bit, 7-bit, and 8-bit dual quantizers, respectively. As the figure reveals, the $(5+1)$-bit quasi-uniform quantizer yields the best FER performance in the error-floor region.

\begin{figure}
\begin{minipage}[b]{0.49\linewidth}
\includegraphics[width=\textwidth]{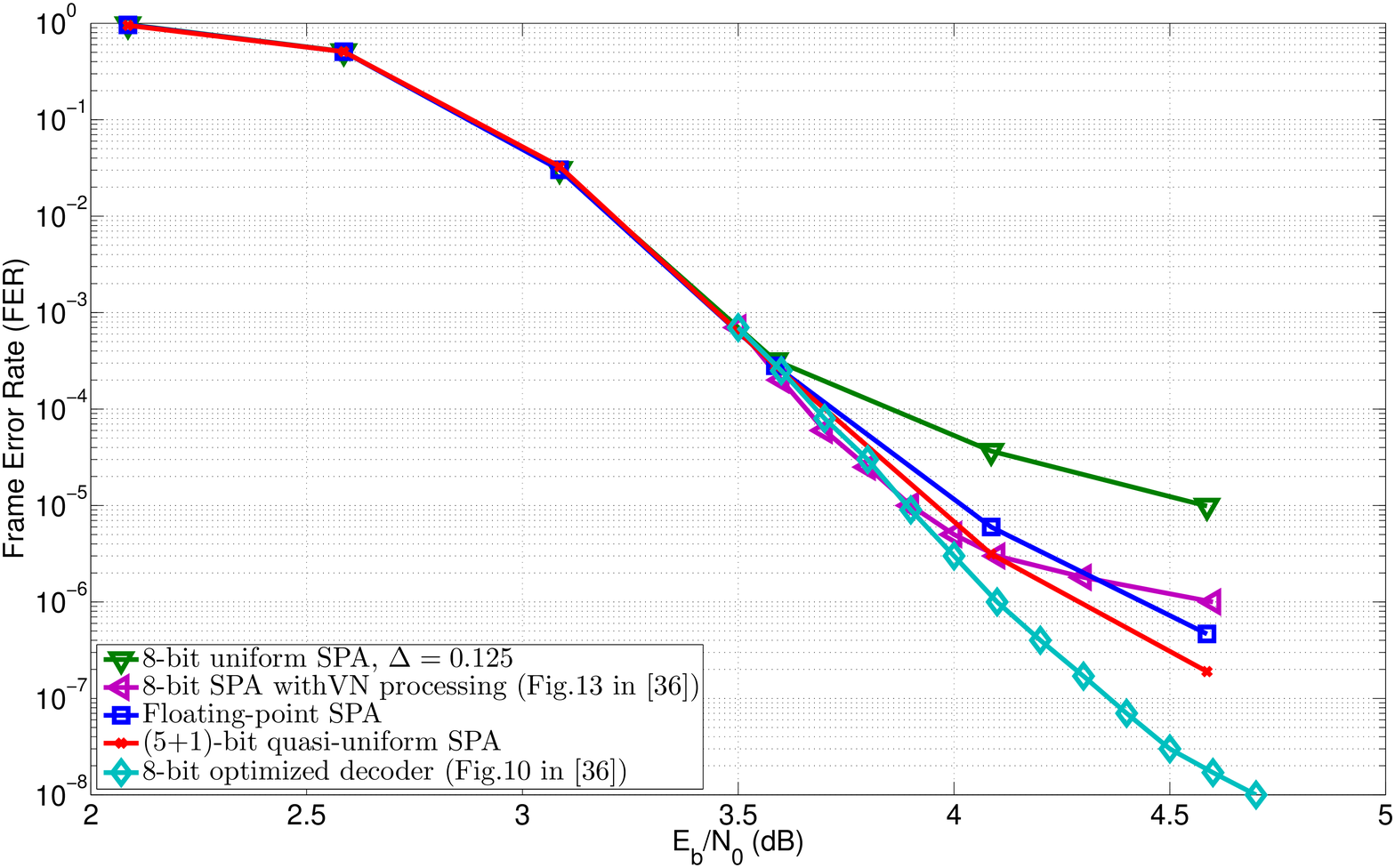}
\centering
\caption{FER results of approx.-SPA decoder on the AR4JA LDPC code of $k=1024$ and $r=0.8$ on AWGNC, where $\Delta=0.5$ and $d=1.3$.}\label{fig_awgn1024spa}
\end{minipage}
\hspace{0.1cm}
\begin{minipage}[b]{0.49\linewidth}
\includegraphics[width=\textwidth]{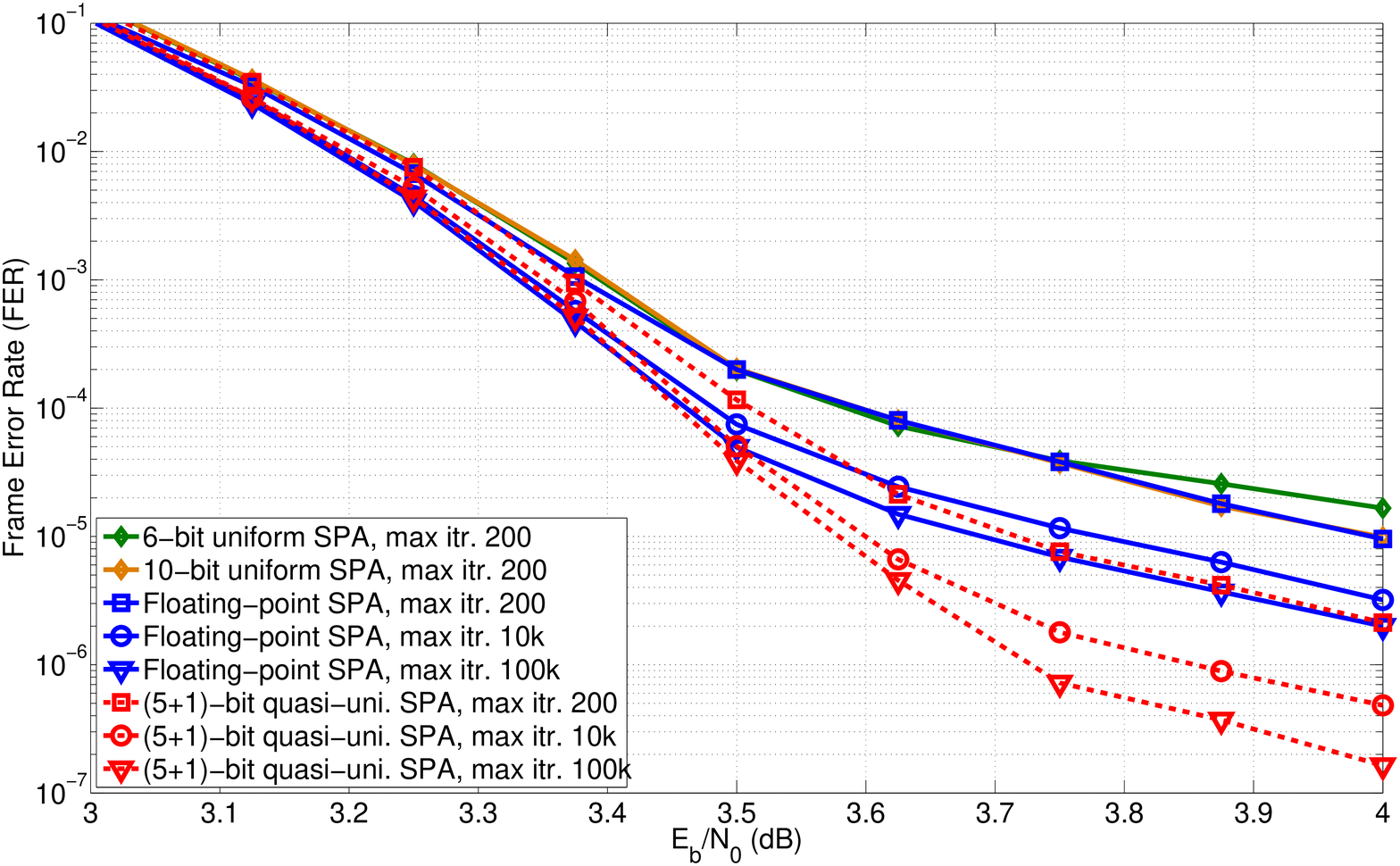}
\centering
\caption{FER results of approx.-SPA decoder on the (4095,3358) LDPC code on AWGNC, where $\Delta=0.5$ and $d=1.3$.}\label{fig_awgn4095spa}
\end{minipage}
\end{figure}

We also evaluated the performance of quasi-uniform quantization in the context of decoding an irregular LDPC code, namely the rate-$\frac{4}{5}$, $(1280,1024)$ AR4JA code. This protograph-based code has variable node degrees ranging from 1 to 6.  Fig.~\ref{fig_awgn1024spa} shows the FER obtained with approximated-SPA decoding and (5+1)-bit quasi-uniform quantization, with $\Delta=0.5$ and  $d=1.3$. Also shown are the results obtained with the floating-point decoder, as well as those produced by 8-bit uniform quantization  with step size $\Delta=0.125$.  The $(5+1)$-bit scheme was superior to both of these alternatives. The figure also includes two curves taken from \cite{Hamkins},
corresponding to an 8-bit quantized SPA decoder with modified VN update rules that were designed specifically for this code, as well as a  ``fully-optimized'' 8-bit decoder with more sophisticated VN/CN update rules. The $(5+1)$-bit quasi-uniform quantizer's performance surpassed that of the former, but it could not match that of the fully-optimized 8-bit decoder.

\subsection{Effect of Iteration Limits}\label{sec:iter_lim}

Figs.~\ref{fig_awgn640ms} -- \ref{fig_awgn1024spa} show that ($q$+1)-bit quasi-uniform quantization can provide attractive error-floor performance, sometimes even better than the double-precision floating-point box-plus SPA decoder. In generating these results, we observed from the simulation data that the floating-point SPA generally requires more iterations to decode a codeword than the quasi-uniform quantized SPA, especially in the high SNR region. Since the maximum number of iterations was set to 200 in our simulations, the faster convergence of the quasi-uniform quantized SPA allowed it to outperform the floating-point SPA scheme. The convergence properties of the quasi-uniform quantized SPA decoder appear to derive from its use of non-uniform, exponentially growing step sizes. From the theoretical analysis discussed in Section~\ref{sec:EF}, we know that the exponential growth rate of correct messages is larger than that of incorrect messages. We might expect that, with  a properly designed  quasi-uniform quantizer, the correct messages can reach the higher magnitude level earlier than the incorrect messages, and therefore incorrect messages are more likely to be quantized to lower magnitude levels. Hence, the correct messages can ``overcome'' the incorrect messages more rapidly, allowing the decoder to converge to a codeword after fewer iterations.

In Fig.~\ref{fig_awgn4095spa}, we explore the effect of limiting the number of iterations in approximated-SPA decoding  of MacKay's rate-0.82, (4095,3358) LDPC code.  With the maximum number of iterations set to 200, we show the results for 6-bit and 10-bit uniform quantizers, the $(5+1)$-bit quasi-uniform quantizer, and the floating-point decoder. We also compare the performance of  $(5+1)$-bit quasi-uniform quantization and the floating-point decoder when the maximum number is raised to 10K and even further to 100K.

With a limit of 200 iterations, this code manifested a high error floor with floating-point SPA decoding. The error floor was lower when the number of iterations could go as high as 10K, and dropped even further when up to 100K iterations were allowed. However, even in the latter case, the FER was only slightly lower than that found with the quasi-uniform quantizer with no more than 200 iterations.
The performance of the quasi-uniform quantizer continued to improve in raising the limit to 10K and then to 100K. These results seem to be consistent with the intuition suggested by the theoretical analysis.

\section{Conclusion}
\label{sec:con}
Trapping sets and other error-prone substructures are known to influence the error-rate performance of LDPC codes with iterative message-passing decoding. In this paper, we have shown that the use of uniform quantization in iterative MP decoding can be a significant factor contributing to the error floor phenomenon in LDPC code performance. An analysis of iterative MP decoding in an idealized setting suggests that decoder message saturation plays a key role in the occurrence of errors in small trapping sets, leading to observed error floor behaviors. To address this problem, we proposed a novel quasi-uniform quantization method that effectively extends the dynamic range of the quantizer. Without modifying the CN and VN update rules or adding extra stages to standard iterative decoding algorithms, the use of this quantizer was shown to significantly lower the error floors of several well-studied LDPC codes when used with various iterative MP decoding algorithms on the BSC and AWGNC. Simulation results confirmed that this new quantization method can significantly reduce the error floors of these codes with essentially no increase in decoding complexity.

\appendices\section{Proof of Theorem~\ref{Thm_MS}}
\label{Appendix_Thm_MS}
\begin{IEEEproof}
Assume VN $v_r\in V_1\subseteq S$ is $k$-separated and the corresponding computation tree is $T(v_r)$.
Let $c_r\in C_1$ be the neighboring degree-one CN of $v_r$ in $S$. From the separation assumption and the assumed correctness of channel messages for VNs outside $S$, all descendants of $c_r$ in $T(v_r)$ receive correct initial messages from the BSC. Like the LLRs of the BSC outputs, all the initial messages in the decoder, $L^{ch}_i$, $1\leq i\leq n$, have the same magnitude. Denote the subtree starting with CN $c_r$ as $T(c_r)$. With the VN/CN update rules of the MS decoder, we analyze the messages sent from the descendants of $c_r$ in $T(c_r)$. First, according to the CN update rule described in \eqref{eq_MS_CN}, all messages received by a VN from its children CNs in $T(c_r)$ must have the same sign as the message received from the channel by this VN, because all the messages passed in $T(c_r)$ are correct.
Therefore, the outgoing message from any VN $v_i$ to its parent CN $c_j$ in $T(c_r)$ satisfies the following equality
\begin{equation}\label{eq_VN_apx}
|L_{i\rightarrow j}| = \left|L^{ch}_i + \sum\limits_{j'\in N(i)\setminus j} L_{j'\rightarrow i}\right|
 = \left|L^{ch}_i\right| + \sum\limits_{j'\in N(i)\setminus j}\left|L_{j'\rightarrow i}\right|.
\end{equation}

Moreover, since the LDPC code considered is variable-regular and all the channel messages from the BSC have the same magnitude,
all incoming messages received by a VN from its children CNs in $T(c_r)$ must have the same magnitude as well. Therefore, all the messages sent from VNs in the same level of the computation tree $T(c_r)$ have the same magnitude.
Let $|L_l|$ be the magnitude of the messages sent by the VNs whose shortest path to a leaf VN contains $l$ CNs in $T(c_r)$; in particular, $|L_0|$ is the magnitude of messages sent by leaf VNs, as well as the magnitude of channel inputs.
The discussion above implies that
\begin{equation}\label{eq_MS_L}
|L_l| = |L_0| + (d_v-1)|L_{l-1}|
> (d_v-1)|L_{l-1}|
> (d_v-1)^l|L_0|
\end{equation}
where $d_v$ is the variable node degree.
Hence, it can be seen that the magnitudes of messages sent towards the root CN $c_r$ of the computation tree $T(c_r)$ grow exponentially, with $d_v-1$ as the base, in every upper VN level.
Therefore, for $l\leq k$, the magnitude of the message sent in the $l$-th iteration from $c_r$ to its parent node $v_r$, the $k$-separated root VN of $T(v_r)$, is greater than $(d_v-1)^l|L_0|$.

Now, let us look at the subtree of $T(v_r)$ that has as its root a child CN $c'\in C_S\setminus C_1$ of the root $v_r$. Denote this subtree by $T(c')$. We assume that the message $L_l'$ received by $v_r$ from $c'$ after $l$ iterations
has a different sign than the message received from $c_r\in C_1$; otherwise, $v_r$ would already have been corrected.
Now consider any
subtree of $T(c')$  that has as its root a VN $v\in S$ and contains $t$ levels of VNs.
We denote such a tree by $T^t(v)$.
If $t\geq 2a$, the subtree $T^t(v)$ must include at least one CN from the set $C_1$.
To see this, recall that the induced subgraph of the trapping set is connected.
Since there are $a$ VNs in the trapping set, it follows that any two VNs in the trapping set can be connected by a path of length less than $2a$.
Therefore, for $t\geq 2a$, $T^t(v)$ actually includes all the CNs and VNs in the induced subgraph of the trapping set, in particular a CN from $C_1$.
Of course, for most trapping sets, $T^t(v)$ can include a CN from $C_1$ with $t$ much smaller than $2a$.

Now,  consider $T^t(v)$ as a ``super-node'' with $(d_v-1)^{t}$ children VNs. Since $T^t(v)$ includes a CN from $C_1$, at least one of these children VNs has the property that all of its descendants receive correct messages from the channel.
This means that at least one of the incorrect messages going into the super-node would be canceled out by one or more such correct messages. 
So if the output message, $L_{out}$, of such a super-node is incorrect, its magnitude satisfies
\begin{equation}\label{eq_MS_Lout}
|L_{out}| < ((d_v-1)^{t}-1)|L_{in}| + |\bar L_{ch}|,
\end{equation}
where $|L_{in}|$ is the largest magnitude of all incoming incorrect messages, and the second term $|\bar L_{ch}|\triangleq|L_0|\sum\limits_{i=0}^{t-1}(d_v-1)^i$ is an upper bound on the sum of the channel input LLRs to all of the VNs in the $t$-level subtree.
Note that the leaf VNs of $T^t(v)$ are not necessarily the leaf VNs of $T(v_r)$.
Thus, we can upper bound the magnitude of the incorrect message sent from $c'$ to $v_r$ after $l$ iterations by
\begin{equation}\label{eq_MS_L'}
\begin{array}{rl}
|L_l'| &< |L_0|\cdot\left[(d_v-1)^{t}-1\right]^{\lceil l / t \rceil} + |\bar L_{ch}|\sum\limits_{i=0}^{\lceil l/t\rceil-1}\left[(d_v-1)^{t}-1\right]^i\\
&<\left(|L_0|+|\bar L_{ch}|\right)\cdot\left[(d_v-1)^{t}-1\right]^{\lceil l / t \rceil}
\end{array}
\end{equation}
where $\lceil x\rceil$ is the smallest integer greater than or equal to $x$. The upper bound in \eqref{eq_MS_L'} is extremely loose, and for most small-size trapping sets, the upper bound is generally less than $|L_0|(d_v-2)^l$.

Therefore, by taking the logarithms of $|L_l|$ in \eqref{eq_MS_L} and $|L'_l|$ in \eqref{eq_MS_L'}, respectively, we have
\begin{equation}\label{eq_logL}
\log|L_l| > \log|L_0| + l\log(d_v-1)
=\log|L_0| + l\cdot\frac{1}{t}\cdot\log(d_v-1)^t,
\end{equation}
and
\begin{eqnarray}\label{eq_logL'}
\log|L_l'| &<& \log\left(|L_0|+|\bar L_{ch}|\right) + \lceil l/t \rceil\log\left[(d_v-1)^{t}-1\right]\\
&<&\log\left(|L_0|+|\bar L_{ch}|\right) + \log\left[(d_v-1)^{t}-1\right] + l\cdot\frac{1}{t}\cdot\log\left[(d_v-1)^{t}-1\right].\nonumber
\end{eqnarray}
Note that the first term in \eqref{eq_logL} and the first two terms in \eqref{eq_logL'} are constants and independent of the number of iterations $l$.

Since $\log(d_v-1)^t > \log\left[(d_v-1)^{t}-1\right]$, if $l$ is large enough and there is no limitation imposed on the magnitude of messages, it is easy to see from \eqref{eq_logL} and \eqref{eq_logL'} that $|L_l|$ would be greater than $|L'_l|$ multiplied by any constant. This means that the correct messages coming from outside of the trapping set to VNs in $V_1$ through their neighboring CNs in $C_1$ will eventually have greater magnitude than the sum of incorrect messages from other neighboring CNs, i.e., $|L_l|>(d_v-1)|L'_l|$. Hence, all the erroneous VNs in $V_1$ will be corrected.
Since, by definition, an absolute trapping set does not contain a stopping set, the remaining erroneous VNs must form a smaller absolute trapping set. Therefore, we can use the same argument to show that as the number of iterations continues to grow, the correct messages would eventually be large enough to correct all erroneous VNs.

Now, we show that the proof technique above can be extended to the AWGNC.
Define $|L_{\min}|$ and $|L_{\max}|$ to be the minimum and maximum magnitudes, respectively,  of the  input LLRs from the AWGNC.
In this setting, the bounds on $\log|L_l|$ and $\log|L_l'|$ corresponding to those in \eqref{eq_logL} and  \eqref{eq_logL'} take the form
\vspace{-0.1in}
\begin{equation}
\label{eq_logL_AWGNC}
\log|L_l| > \log|L_{\min}| + l\cdot\frac{1}{t}\cdot\log(d_v-1)^t,
\vspace{-0.1in}
\end{equation}
and
\begin{equation}
\label{eq_logL'_AWGNC}
\log|L_l'|<\log\left(|L_{\max}| + |\bar L_{ch}| \right) + \log\left[(d_v-1)^{t}-1\right]
 + l\cdot\frac{1}{t}\cdot\log\left[(d_v-1)^{t}-1\right].
\vspace{-0.1in}
\end{equation}
Since the quantities $\log|L_{\min}|$ and $\log|L_{\max}|$ are  constant and do not change as $l$ increases,  we can conclude, as we did for the BSC, that the correct messages from outside the trapping set will eventually have greater magnitude than the incorrect messages from within the trapping set. Therefore, all of the VNs will eventually be correctly decoded.
\end{IEEEproof}

\section{Proof of Corollary~\ref{Cor_MMS}}
\label{Appendix_Cor_MMS}
\begin{IEEEproof}
We first consider AMS decoding. Referring to the proof of Theorem~\ref{Thm_MS} for the BSC case, we can replace the quantity $(d_v-1)$ in \eqref{eq_logL} and \eqref{eq_logL'} by $\alpha(d_v-1)$, where $\alpha$ is the attenuation factor.
In practice, we would always choose $\alpha$ such that $\alpha(d_v-1)$ is greater than 1; otherwise, the error-correction performance of the AMS decoder would be inferior to that of the MS decoder. Similar reasoning to that used in the proof of the theorem then leads to the desired conclusion.  For the AWGNC case, we make the corresponding changes in \eqref{eq_logL_AWGNC} and \eqref{eq_logL'_AWGNC}, and argue similarly.


For the OMS decoder, the proof follows from the proof of  Theorem~\ref{Thm_SPA} in Appendix~\ref{Appendix_Thm_SPA}. There, we simply replace the quantity $\bar s$ by the offset $\beta$.
\end{IEEEproof}

\section{Proof of Lemma~\ref{Lemma_MS_vs_SPA}}
\label{Appendix_Lemma_MS_vs_SPA}
\begin{IEEEproof}
The first statement regarding the relationship between the sign and magnitude of the CN messages $L_{SPA}$ and $L_{MS}$ is proved in \cite{Fossorier},\cite{MinSum_vari}. For completeness, we include here an elementary alternative proof.

First note that if $xy=0$, then $s(x,y)=0$. Now, if $x$ and $y$ are nonzero and have the same sign (i.e., $xy>0$), then $|x+y|>|x-y|$ and hence $s(x,y)<0$. Hence, we can see from \eqref{eq_boxplus_m} that the first statement is true if the inequality $\min(x,y) + s(x,y) > 0 $ holds for any positive real values $x$ and $y$. Without loss of generality, if we assume $x\geq y>0$, then the following inequalities are equivalent
\begin{equation}\notag
\begin{array}{rrl}
&\min(x,y) + s(x,y) >& 0\\
\Leftrightarrow& \log{e^{y}} + \log\frac{1+e^{-x+y}}{1+e^{-x-y}} > &0 \\
\Leftrightarrow& e^{y} + e^{-x} -1 - e^{-x+y} > &0 \\
\Leftrightarrow& (e^{y}-1) (1- e^{-x}) >& 0.
\end{array}
\end{equation}
Since $e^{y} > 1$ and $e^{-x} < 1$, the final inequality holds. Hence, the first statement is proved.

To prove the second statement, note that

\begin{equation}\notag
s(x,y) = \log\frac{1 + e^{-x-y}}{1 + e^{-x+y}}
=\log\frac{e^x + e^{-y}}{e^x + e^y}
\geq \log\frac{e^x + e^{-x}}{e^x + e^x} > \log\frac{1}{2} = -\log2.
\end{equation}
Therefore,  $-\log2<s(x,y)<0$. When $xy<0$, a similar line of reasoning shows that $s(x,y)>0$ and $0<s(x,y)<\log2$.
\end{IEEEproof}
\section{Proof of Theorem~\ref{Thm_SPA}}
\label{Appendix_Thm_SPA}
\begin{IEEEproof}
From Lemma~\ref{Lemma_MS_vs_SPA}, we know that a CN message in SPA decoding has the same sign as the corresponding CN message in MS decoding. Moreover, the magnitude of the former is  less than or equal to that of the latter.
To compute the output for a CN of degree $d_c$, the box-plus SPA uses the pairwise box-plus operation \eqref{eq_boxplus_p} at most $\log (d_c-1)$ times. Hence, the difference between output messages of the SPA and the MS algorithm is upper bounded by $\bar s \triangleq \lceil\log (d_c-1)\rceil\cdot\log2$.

By applying an approach similar to that used in the proof of Theorem~\ref{Thm_MS}, we can lower bound the magnitude of messages $L_l$ in SPA decoding as follows
\begin{eqnarray}
\label{eq_SPA_L}
|L_l| &>& |L_0| + (d_v-1)\left(|L_{l-1}| - {\bar s}\right)\nonumber\\
&>& (d_v-1)^l|L_0|-{\bar s}\sum\limits_{i=1}^{l}{(d_v-1)^i}\nonumber\\
&=& (d_v-1)^l\left(|L_0|-\frac{d_v-1}{d_v-2}{\bar s}\right) + {\bar s} \frac{d_v-1}{d_v-2}.\nonumber
\end{eqnarray}

Since all input messages to the decoder from the BSC have the same magnitude, if we scale the magnitudes of all initial messages such that
\begin{equation}
\label{eq_L0}
|L_0|>\frac{d_v-1}{d_v-2}\bar s
=\frac{d_v-1}{d_v-2}\cdot\lceil\log(d_c-1)\rceil\cdot\log2,
\end{equation}
then the magnitudes of messages sent towards $c_r$ in the computation tree $T(v_r)$ grow exponentially in the number  of iterations, with base $d_v-1$.
Hence, using the same reasoning as in the proof of Theorem~\ref{Thm_MS}, it can be shown that, if $k$ is large enough and there is no limit on the magnitudes of messages, the correct messages from outside the trapping set eventually overcome the incorrect messages passed within the trapping set, thereby correcting all erroneous VNs in the trapping set.

The extension to the AWGNC case is analogous to that used in Theorem~\ref{Thm_MS}. Let $|L_0|$ denote the minimum magnitude of all input LLRs from the AWGNC, and linearly scale the magnitudes of all the input messages such that the inequality \eqref{eq_L0} is satisfied. Then, reasoning as in the proof of the BSC case above, we can show that the magnitudes of correct messages outside the trapping set still grow exponentially with $d_v-1$ as the base, and eventually they correct all erroneous VNs in the trapping set.
\end{IEEEproof}

\section*{Acknowledgment}
The authors would like to thank Yang Han and William Ryan for providing the parity check matrix of the (640,192) QC LDPC code, Brian Butler for helpful discussions, and the anonymous reviewers for their numerous and detailed suggestions that helped to improve this paper.

\ifCLASSOPTIONcaptionsoff
  \newpage
\fi

\end{document}